\documentclass[prd, superscriptaddress, tightenlines, longbibliography, nofootinbib, eqsecnum, amsfonts, amsmath, floatfix, twocolumn, notitlepage]{revtex4-2}
\pdfoutput=1

\usepackage[utf8]{inputenc}
\usepackage{mathrsfs}
\usepackage{euscript}
\usepackage{graphics}
\usepackage{graphicx}
\usepackage{amsmath}
\usepackage{amssymb}
\usepackage{gensymb}
\usepackage{bm}
\usepackage[usenames,dvipsnames,svgnames,table]{xcolor}
\definecolor{linkcolor}{rgb}{0.6,0,0}
\definecolor{citecolor}{rgb}{0,0,0.75}
\definecolor{urlcolor}{rgb}{0.12,0.46,0.7}
\usepackage{xspace}
\usepackage{wasysym}
\usepackage{times}
\usepackage{appendix}
\usepackage{comment}
\usepackage{lipsum}
\usepackage[nolist,nohyperlinks]{acronym}
\usepackage{float}
\usepackage{simplewick}
\usepackage{natbib, ifthen}
\usepackage[breaklinks, colorlinks, urlcolor=urlcolor, linkcolor=linkcolor,citecolor=citecolor,pdfencoding=auto]{hyperref}
\usepackage{longtable}
\usepackage{listings}

\newcommand{\la}{\langle}

\newcommand{\mksym}[1]{\ifmmode {\rm #1}\else #1\fi}

\providecommand{\lea}{\la}

\providecommand{\alt}{\lea}

\providecommand{\text}[1]{\rm{#1}}

\newcommand{\begm}{\begin{pmatrix}}
\newcommand{\enm}{\end{pmatrix}}

\newcommand\ba{\begin{eqnarray}}
\newcommand\ea{\end{eqnarray}}
\newcommand\bea{\begin{eqnarray}}
\newcommand\eea{\end{eqnarray}}

\newcommand\be{\begin{equation}}
\newcommand\ee{\end{equation}}

\newcommand{\boldvec}[1]{{\mbox{\boldmath{$#1$}}}}

\newcommand{\vL}{\boldvec{L}}

\newcommand{\vl}{\boldvec{l}}

\defcitealias{Aghanim:2018oex}{PL2018}

\defcitealias{Pratten:2016dsm}{PB16}

\newcommand{\isdraft}[1]{}

\begin{document}
\title{How to detect lensing rotation}

\newcommand{\Sussex}{Department of Physics \& Astronomy, University of Sussex, Brighton BN1 9QH, UK}

\author{Mathew Robertson}
\affiliation{\Sussex}

\author{Antony Lewis}
\affiliation{\Sussex}

  \begin{abstract}
    Gravitational lensing rotation of images is predicted to be negligible at linear order in density perturbations, but can be produced by the post-Born lens-lens coupling at second order. This rotation is somewhat enhanced for Cosmic Microwave Background (CMB) lensing due to the large source path length, but remains small and very challenging to detect directly by CMB lensing reconstruction alone.
We show the rotation may be detectable at high significance as a cross-correlation signal between the curl reconstructed with Simons Observatory (SO) or CMB-S4 data, and a template constructed from quadratic combinations of large-scale structure (LSS) tracers.
    Equivalently, the lensing rotation-tracer-tracer bispectrum can also be detected, where LSS tracers considered include the CMB lensing convergence, galaxy density, and the Cosmic Infrared Background (CIB), or optimal combinations thereof.
We forecast that an optimal combination of these tracers can probe post-Born rotation at the level of $5.7\sigma$--$6.1\sigma$ with SO and $13.6\sigma$--$14.7\sigma$ for CMB-S4, depending on whether standard quadratic estimators or maximum a posteriori iterative methods are deployed. We also show possible improvement up to $21.3\sigma$ using a CMB-S4 deep patch observation with polarization-only iterative lensing reconstruction.
However, these cross-correlation signals have non-zero bias because the rotation template is quadratic in the tracers, and exists even if the lensing is rotation free. We estimate this bias analytically, and test it using simple null-hypothesis simulations to confirm that the bias remains subdominant to the rotation signal of interest. Detection and then measurement of the lensing rotation cross-spectrum is therefore a realistic target for future observations.

  \end{abstract}


   \maketitle

\section{Introduction}

Gravitational lensing of Cosmic Microwave Background (CMB) photons by the matter along our line of sight distorts the observed CMB fields, giving a distinct observational signature of lensing (see Ref.~\cite{Lewis:2006fu} for a review).
The physics generating the primordial anisotropies in the CMB fields are well understood, and they are very close to Gaussian, so the non-Gaussian statistics of the observed lensed fields contain enough information to reconstruct the lensing field up to statistical fluctuations. The power spectrum of the lensing field depends on the large-scale structure (LSS) density fluctuations and late-time geometry of the Universe, and hence provides an observational probe of the Universe after recombination.
The CMB lensing convergence has now been detected to high significance by multiple experiments \cite{PL2018,Wu:2019hek,Sherwin:2016tyf,BICEP2:2016rpt,POLARBEAR:2019snn,Carron:2022eyg, ACT:2023dou},
corresponding to purely gradient lensing deflection angles as predicted for first-order lensing by density perturbations.

Higher-order corrections to the lensing field can however produce lensing rotation from scalar perturbations, as predicted by a second-order post-Born lensing expansion~\cite[]{Jain:1999ir,Cooray:2002mj,Krause:2009yr,Pratten:2016dsm,Beck:2018wud}. This corresponds to having a curl-like component to the lensing deflection field~\cite[]{Cooray:2005hm}, and
is produced by a lens-lens coupling between non-aligned shear distortions induced by two gravitational potentials at different redshifts. Since the rotation is second order, it is predicted to be much smaller than the convergence signal. For current experiments it can be taken to be negligible, so that the measured rotation signal can serve as a null test. The predicted lensing post-Born rotation can in principle be measured by lensing quadratic estimators with more sensitive future data; however, as we shall describe, even using iterative estimators with fourth-generation (CMB-S4) experiments, it is challenging to detect directly at any significance due to the large lensing reconstruction noise.

Ref.~\cite{Pratten:2016dsm} showed that the rotation may instead be detectable with CMB-S4 via the convergence-convergence-rotation bispectrum, though more futuristic data would still be required to get much above a $3 \sigma$ detection. Conceptually, this bispectrum estimator is using two convergence modes to probe the two lenses that combine to produce the rotation via lens-lens coupling, giving an approximate quadratic estimate of the lensing rotation. The cross-correlation of this quadratic rotation estimate with the measured rotation then gives a probe of the lensing rotation power. This immediately suggests that we could do much better: by using other tracers of the large-scale structure with lower noise and better redshift resolution, we can trace the multiple redshifts of the lens-lens couplings to produce a more accurate rotation template. In the limit in which we had a perfect tracer of the LSS mass distribution, it would be possible to predict exactly the lensing rotation that would be produced by ray tracing through it. This perfect rotation template could then be correlated with the lensing reconstruction rotation field, giving a much more sensitive probe of the rotation than the reconstruction auto-spectrum. In this paper we investigate how close we can get to this ideal situation, by combining more realistic large-scale probes (such as galaxy counts and Cosmic Infrared Background (CIB)) as well as the CMB lensing convergence.

We begin with an overview of weak lensing observables and modelling of their power spectra, including post-Born rotation, in Section~\ref{sec:background}. We then examine the difficulty of direct lensing rotation reconstruction in Section~\ref{sec:init_fisher}, but prove that detection is feasible for near-future CMB missions with some appropriately constructed rotation template. In Section~\ref{sec:template} we construct an explicit estimator for this lensing rotation template, and relate its cross-spectrum with the measured rotation to the corresponding bispectrum estimators. We then make forecasts in Section~\ref{sec:fisher} for the upcoming Simons Observatory (SO) \cite[]{Ade:2018sbj} experiment and also for the next generation ``stage 4'' (CMB-S4) \cite[]{Abazajian:2019eic} experiment, combined with various LSS tracers, showing that the rotation signal should be detectable at high significance. As a first-look analysis, we make fairly idealized assumptions, but show that even with idealized data there will be some non-zero biases affecting the rotation power spectrum estimate. Section~\ref{sec:bias} describes the origin of these biases, and we construct an explicit analytic model which we then test against simulations. Our new estimators and models provide a way to clearly detect the lensing rotation signal with future data, and can also be used to quantify the expected signal in mixed rotation bispectra that may be useful null tests for other cosmological analyses. We finish by summarizing such conclusions in Section~\ref{sec:conclusion}.

Throughout this paper we assume a standard flat $\Lambda$CDM cosmology consistent with Planck \cite[]{Ade:2015xua}. This only enters at the level of the matter power spectrum computed by {\sc CAMB} \cite[]{Lewis:1999bs}, with non-linear corrections modelled by {\sc HMCode} \cite{Mead:2015yca}.

\section{Background}\label{sec:background}
Weak lensing is the statistical distortion of an observation due to the gravitational influence of the intervening LSS.  Photons are deflected from their background geodesic path by an angle proportional to the gradient of the gravitational potential $\boldsymbol{\nabla}\Psi(\chi)$.  The total deflection, $\boldsymbol{\alpha}$, is the accumulation of all gravitational effects inflicted upon a particular ray \cite[]{Lewis:2006fu}
\begin{equation}\label{eq:alpha_1}
\boldsymbol{\alpha}(\boldsymbol{\hat{n}})=-2\int^{\chi_s}_0 d\chi W(\chi, \chi_s)\boldsymbol{\nabla}\Psi(\boldsymbol{\hat{n}},\chi),
\end{equation}
where the window function, $W(\chi, \chi_s)$, is the lensing kernel for an image at source distance $\chi_s$,  and $\chi$ is the comoving radial distance.

The observed position, $\boldsymbol{\hat{n}}(\theta,\phi)$, of an observable, X, on the celestial sphere has been transported from its true position by a deflection angle $\boldsymbol{\alpha}$, so that the measured {\it lensed} observable is actually a remapping of the unlensed field\footnote{It is not exactly a remapping, but remapping is a good approximation~\cite{Lewis:2017ans}.},
\begin{equation}
	\tilde{X}(\boldsymbol{\hat{n}}) = X(\boldsymbol{\hat{n}}+\boldsymbol{\alpha}(\boldsymbol{\hat{n}})).
\end{equation}

In Eq.~\eqref{eq:alpha_1}, at lowest order $\boldsymbol{\alpha}$ is determined by integrating the contributions of $\boldsymbol{\nabla}\Psi$ along the background geodesic.  However,  photons do not travel along background geodesics due to the lensing deflection. To account for these deviations, perturbative corrections to the angular position at which the gravitational potential is evaluated along the radial path must be applied within the integral. At some arbitrary point along the radial path at position $\hat{\boldsymbol{n}}_0$, we can expand about the background geodesic to second order \cite[]{Pratten:2016dsm}
\begin{multline}
    \begin{split}
	\Psi(\boldsymbol{\hat{n}})&=\Psi(\boldsymbol{\hat{n}}_0+\boldsymbol{\alpha}(\boldsymbol{\hat{n}}_0))\\
    &= \Psi(\boldsymbol{\hat{n}}_0)+\left[\alpha^a\nabla_a\Psi\right](\boldsymbol{\hat{n}}_0) +\mathcal{O}(\Psi^3),
    \end{split}
\end{multline}
where implicit dependence between the gravitational potential and deflection angle results in the iterative solution to Eq.~\eqref{eq:alpha_1}, which to second order is
\begin{multline}\label{eq:alpha_pert}
\alpha_a(\boldsymbol{\hat{n}}_0)= -2\int^{\chi_s}_0d\chi W(\chi,\chi_s)\Bigg[\nabla_a\Psi(\chi)\\
-2\int^{\chi}_0d\chi' W(\chi',\chi)\nabla_a\nabla_b\Psi(\chi)\nabla^b\Psi(\chi')
\Bigg](\boldsymbol{\hat{n}}_0)\\
+\mathcal{O}(\Psi^3).
\end{multline}

Higher order corrections to $\boldsymbol{\alpha}$ are relatively tame to compute, but here we are only interested in the leading-order contribution to the lensing rotation, which sits nicely at second order.

\subsection{Lensing Convergence}\label{sec:conv}
It can be seen from Eq.~\eqref{eq:alpha_pert} (or Eq.~\eqref{eq:alpha_1}) that to first order the total deflection is determined by the gravitational potential along the background geodesic.  This is known as the Born approximation, which describes one effective total deflection from the original path of a photon. At this order the deflection field is curl free\footnote{As we are working on the two-dimensional sky, the cross product here is defined as $(\boldsymbol{\nabla}\times A)_a=\epsilon_{ab}\nabla_bA$ where $\boldsymbol{\epsilon}$ is the antisymmetric tensor or the 2-dimensional Levi-Cevita symbol.}, i.e. $\boldsymbol{\nabla}\times\boldsymbol{\alpha}=0$, and thus is fully described as the angular gradient of some function
\begin{equation}\label{eq:alpha}
	\boldsymbol{\alpha}(\boldsymbol{\hat{n}})=\boldsymbol{\nabla}\phi(\boldsymbol{\hat{n}}),
\end{equation}
where that function $\phi$ is known as the {\it lensing potential}. At this order $\phi$ is simply the projection of the gravitational potentials sourced from density perturbations along the unperturbed geodesic.

The {\it lensing convergence} is the observable describing the magnification effect and is directly related to the lensing potential by
\begin{equation}
	\kappa(\boldsymbol{\hat{n}})=-\frac{1}{2}\boldsymbol{\nabla}\cdot \boldsymbol{\alpha}(\boldsymbol{\hat{n}})=-\frac{1}{2}\boldsymbol{\nabla}^2\phi(\boldsymbol{\hat{n}}).
\end{equation}
Working within the flat sky approximation of the two dimensional spherical sky, the convergence in harmonic space is written
\begin{equation}\label{eq:kappa_harm}
	\kappa(\boldsymbol{L})=-L^2\int^{\chi_s}_0d\chi W(\chi,\chi_s) \Psi(\boldsymbol{L},\chi).
\end{equation}
For CMB lensing, the source is emitted at the surface of last scattering, $\chi_s=\chi_{*}$, which can be taken as a single source plane to high accuracy.

\subsubsection{Convergence power spectrum}
Using the usual definition for the power spectrum of two fields in harmonic space $\langle X(\boldsymbol{L})Y(\boldsymbol{L}')\rangle=(2\pi)^2\delta_{\textrm{D}}(\boldsymbol{L}+\boldsymbol{L}')C^{XY}_{L}$, and the lowest-order Limber approximation, $k\approx L/\chi$~\cite[]{Kaiser:1991qi}, the convergence power spectrum to leading order is
\begin{multline}\label{eq:kappa_ps}
    \begin{split}
	C^{\kappa\kappa}_L(\chi_1,\chi_2)=\int^{\chi_2}_0d\chi\frac{W_{\kappa}(\chi,\chi_1)W_{\kappa}(\chi,\chi_2)}{\chi^2}\\
    P_{\delta\delta}\left(k\approx\frac{L}{\chi},z(\chi)\right),
        \end{split}
\end{multline}
for two sources at radial distances $\chi_1$ and $\chi_2$ where $\chi_2\geq\chi_1$. This distinction between source planes becomes necessary later in the paper when defining lensing bispectra.

Note that we have switched from integrating over the Weyl potential, $\Psi$, to the comoving-gauge matter density perturbation, $\delta$, for later notational convenience. The conversion is achieved through the cosmological Poisson equation which has a simple form during matter and dark energy domination (assuming General Relativity and neglecting small effects from massive neutrinos):
\begin{equation}\label{eq:poisson}
	k^2 \Psi(\boldsymbol{k}, \chi)\approx-\gamma(\chi) \delta(\boldsymbol{k},\chi)=-\frac{3\Omega_m H_0^2}{2a(\chi)}\delta(\boldsymbol{k},\chi),
\end{equation}
for scale factor $a(\chi)=\left(1+z(\chi)\right)^{-1}$,  present-day density parameter of matter, $\Omega_m$, and Hubble constant, $H_0$.

The window function for the CMB lensing convergence is\footnote{For the CMB lensing convergence, $\chi_s=\chi_*$. However we leave the kernel in this more general form as it is useful for the definition of the curl power spectrum Eq.~\eqref{eq:omega_ps}, and the galaxy lensing window function Eq.~\eqref{eq:window_shear}.}
\begin{equation}\label{eq:window_cmb}
	W_{\kappa}(\chi,\chi_s)=\gamma(\chi)\chi^2\frac{\chi_s-\chi}{\chi\chi_s}\Theta(\chi_s-\chi),
\end{equation}
where the Heaviside function $\Theta$ ensures $\chi_s\geq\chi$. For weak lensing with galaxies, the galaxy sources lie within many redshift shells, unlike the CMB which is a single source plane to good approximation. Taking this into consideration, the lensing kernel for galaxy lensing radially integrates over the source distribution, giving

\begin{equation}\label{eq:window_shear}
	W_{\kappa^{\textrm{gal}}}(\chi)=\int_{\chi}^{\infty} d\chi' n(z')H(z')W_{\kappa}(\chi,\chi'),
\end{equation}
where the distribution of source galaxies, $n$, is given as function of redshift, $z$, and is thus accompanied by the Hubble parameter, $H(z)$, as the conversion factor to $\chi$.

Throughout this paper, we will only be using the lensing convergence within the Born approximation. The leading post-Born corrections to the convergence power spectrum are at the $0.2\%$ level \cite[]{Pratten:2016dsm, Krause:2009yr} and are negligible for our analysis.

The accuracy of the convergence power spectrum is significantly improved at large scales using the extended Limber approximation, $k\approx(L+1/2)/\chi$ \cite{LoVerde:2008re}. However, we use the lowest order Limber approximation for modelling of second order post-Born statistics, such as rotation (which we will describe in terms of the convergence power spectrum), hence for consistency we keep Eq.~\eqref{eq:kappa_ps} to lowest order.

\subsection{Post-Born Rotation}\label{sec:omega}
To get field rotation we need to go beyond the Born approximation and into the post-Born regime. Looking at the second order term in Eq.~\eqref{eq:alpha_pert} we see that the deflection field is no longer curl free, and thus cannot be fully described solely by the lensing potential. We introduce the pseudo-scalar {\it curl potential} $\Omega$ \cite[]{Cooray:2005hm} so that
\begin{equation}\label{eq:alpha_rot}
	\boldsymbol{\alpha}(\boldsymbol{\hat{n}})=\boldsymbol{\nabla}\phi(\boldsymbol{\hat{n}}) + \boldsymbol{\nabla}\times\Omega(\boldsymbol{\hat{n}}).
\end{equation}
Analogous to the lensing convergence, the curl field is accessible through the deflection field via the {\it lensing rotation} observable
\begin{equation}
	\omega(\boldsymbol{\hat{n}})=-\frac{1}{2}\epsilon^{ab}\nabla_b\alpha_a(\boldsymbol{\hat{n}})=-\frac{1}{2}\boldsymbol{\nabla}^2\Omega(\boldsymbol{\hat{n}}).
\end{equation}
This describes the rotational distortion to a field, and first appears at second order due to the {\it lens-lens} interaction of a particular photon bundle. That is, the result of two sequential lensing events (specifically, two unaligned shearings) induces rotation. The explicit form of field rotation in harmonic space (again in the flat sky limit) is
\begin{equation}
\begin{split}\label{eq:omega}
	\omega(\boldsymbol{L})=&-2\int^{\chi_s}_0d\chi W(\chi,\chi_s)\int^{\chi}_0d\chi' W(\chi',\chi)\\
	&\times\int\frac{d^2\boldsymbol{l}}{(2\pi)^2}(\boldsymbol{l}\cdot\boldsymbol{l}')[\boldsymbol{l}\times\boldsymbol{L}] \Psi(\boldsymbol{l},\chi)\Psi(\boldsymbol{l}',\chi'),
\end{split}
\end{equation}
where $\boldsymbol{l}'\equiv\boldsymbol{L}-\boldsymbol{l}$ for brevity.

Another second order effect comes from the {\it ray-lens} interaction. This correction is not due to a second lensing event, but instead to a change in the gravitational potential gradients along the deflected path of the photon.  However, it produces no rotation, and we are not considering second order corrections to the convergence here.

\subsubsection{Rotation power spectrum}
The rotation power spectrum is \cite[]{Pratten:2016dsm}
\begin{equation}\label{eq:omega_ps}
    C^{\omega\omega}_{L}(\chi_s)=4\int \frac{d^2\boldsymbol{l}}{(2\pi)^2}\frac{
    \left(\boldsymbol{l}\cdot\boldsymbol{l}'\right)^2\left[\boldsymbol{l}\times\boldsymbol{L}\right]^2}{l^4\left(l'\right)^4}
    \mathcal{M}^{\omega\omega}_{\chi_s}\left(l,l'\right),
\end{equation}
where the rotation mode-coupling matrix is defined as
\begin{multline}
    \begin{split}\label{eq:omega_mc}
    \mathcal{M}^{\omega\omega}_{\chi_s}(l,l')=&\int^{\chi_s}_{0}d\chi \frac{W_{\kappa}(\chi,\chi_s)^2}{\chi^2}\\
    &P_{\delta\delta}\left(k\approx\frac{l}{\chi},z(\chi)\right)C^{\kappa\kappa}_{l'}(\chi,\chi).
        \end{split}
\end{multline}
We only consider the auto spectrum of lensing rotation estimators at the same redshift, as appropriate for the CMB, so $C^{\omega\omega}_{L}$ only depends on a single source plane.

For a more detailed description of the post-Born derivation and in-depth discussion of its impact, see Ref.~\cite[]{Pratten:2016dsm}.

\section{Is Rotation Detectable?}\label{sec:init_fisher}
The rotation signal from post-Born lensing is a second order lensing effect, so it is subdominant to convergence and shear from the gradient deflection field.  We now explore whether the rotation is actually detectable within the near future using current lensing estimator technology.

\subsection{CMB Lensing Reconstruction}\label{sec:cmb_lensing}
A fixed weak lensing field breaks the statistical isotropy of the CMB sky, introducing correlations between unequal modes \cite[]{Lewis:2006fu}. These new correlations can be probed to extract information about the lensing field that produced them. This is typically done with quadratic estimators \cite[]{Hu:2001kj, Okamoto:2003zw} for their computational efficiency. In Ref.~\cite[]{Maniyar:2021msb}, a global minimum variance (GMV) quadratic estimator (QE) for the lensing potential is constructed using quadratic combinations of CMB fields weighted by the inverse of their covariance. For lensed CMB fields $\tilde{X}\in \{\tilde{T},\tilde{E},\tilde{B}\}$, one can similarly construct a GMV estimator for the field rotation
\begin{equation}\label{eq:omega_qe}
	\hat{\omega}(\boldsymbol{L})=\frac{L^2}{2}A^{\Omega}_L\int\frac{d^2\boldsymbol{l}}{(2\pi)^2}\tilde{X}^i(\boldsymbol{l})g^{\Omega}_{ij}(\boldsymbol{l},\boldsymbol{L})\tilde{X}^j(\boldsymbol{l}'),
\end{equation}
with corresponding inverse covariance weighted functions
\begin{equation}\label{eq:gmv_weight}
    g^{\Omega}_{ij}(\boldsymbol{l},\boldsymbol{L})=\frac{1}{2}(\tilde{\boldsymbol{C}}_\textrm{CMB}^{-1})^{ip}_{l}f^{\Omega}_{pq}(\boldsymbol{l},\boldsymbol{L})(\tilde{\boldsymbol{C}}_{\textrm{CMB}}^{-1})^{jq}_{l'}.
\end{equation}
The response functions, $f^{\Omega}_{ij}(\boldsymbol{l},\boldsymbol{L})$, quantify the correlations between modes $\boldsymbol{l}$ and $\boldsymbol{l}'$ that are induced by the curl potential,
\begin{equation}\label{eq:resp}
\left\langle\frac{\delta}{\delta\Omega(\boldsymbol{L})}\tilde{X}(\boldsymbol{l})\tilde{Y}(\boldsymbol{l}')\right\rangle=\delta(\boldsymbol{l}+\boldsymbol{l'}-\boldsymbol{L})f^{\Omega}_{XY}(\boldsymbol{l},\boldsymbol{L}),
\end{equation}
and are explicitly stated in Table \ref{tab:resps}. The covariance matrices include the lensed CMB spectra plus noise, $(\tilde{\boldsymbol{C}}_\textrm{CMB})^{ij}_l=\tilde{C}^{ij}_{l}+N^{ij}_{l}$. For simplicity we only consider complete flat-sky observations with isotropic noise, so that the covariances are diagonal in $\boldsymbol{l}$.

\begin{table}[]\centering
\begin{tabular}{|l|l|}
    \hline
 $XY$ & $f^{\Omega}_{XY}(\boldsymbol{l},\boldsymbol{L})$\\ \hline \hline
$TT$ & $\boldsymbol{l}\times\boldsymbol{L}\left(\tilde{C}^{T\nabla T}_{l} - \tilde{C}^{T\nabla T}_{l'}\right)$\\
$EE$ & $\boldsymbol{l}\times\boldsymbol{L}\left(\tilde{C}^{E\nabla E}_{l} - \tilde{C}^{E\nabla E}_{l'}\right)h_E(\boldsymbol{l},\boldsymbol{l}')$\\
$EB$ & $\boldsymbol{l}\times\boldsymbol{L}\left(\tilde{C}^{E\nabla E}_{l} - \tilde{C}^{B\nabla B}_{l'}\right)h_B(\boldsymbol{l},\boldsymbol{l}')$\\
$TE$ & $\boldsymbol{l}\times\boldsymbol{L}\left(\tilde{C}^{T\nabla E}_{l}h_E(\boldsymbol{l},\boldsymbol{l}') - \tilde{C}^{T\nabla E}_{l'}\right)$\\
$TB$ & $\boldsymbol{l}\times\boldsymbol{L}\tilde{C}^{T\nabla E}_{l}h_B(\boldsymbol{l},\boldsymbol{l}')$\\ \hline
\end{tabular}
    \caption{\small{The curl potential response functions for individual pairs of CMB maps $X$ and $Y$ required for lensing reconstruction. The geometric factors are defined as $h_E(\boldsymbol{l_1},\boldsymbol{l_2})=\cos(2(\theta_{l_1}-\theta_{l_2}))$, and $h_B(\boldsymbol{l_1},\boldsymbol{l_2})=\sin(2(\theta_{l_1}-\theta_{l_2}))$, and $\boldsymbol{l}'=\boldsymbol{L}-\boldsymbol{l}$. The non-perturbative response functions given here (slightly generalizing previous results \cite{Hu:2001kj,Cooray:2005hm}) use the lensed gradient spectra defined in Ref.~\cite[]{Lewis:2011fk}. The BB response function is excluded along with the curl-like terms, $\tilde{C}^{XP_{\perp}}$, as they are subdominant relative to the other terms \cite[]{Hanson:2010rp}.}}\label{tab:resps}
\end{table}

The quadratic estimator is normalized by
\begin{equation}\label{eq:N0}
    [A^{\Omega}_L]^{-1}=\left[N_0^{\Omega}(L)\right]^{-1}=\int \frac{d^2\boldsymbol{l}}{(2\pi)^2}g_{ij}^{\Omega}(\boldsymbol{l},\boldsymbol{L})f^{\Omega}_{ij}(\boldsymbol{l},\boldsymbol{L}),
\end{equation}
which is conveniently related to the lowest order reconstruction noise, $N_0^{\omega}(L)=L^4N_0^{\Omega}(L)/4$, on the reconstructed field rotation. The GMV estimator for the CMB lensing convergence is very similar, only differing slightly in the response functions (which can be found in Ref.~\cite[]{Maniyar:2021msb}).

For low noise levels the quadratic estimator becomes suboptimal, effectively being limited by the lensing variance in the lensed B modes. Iterative reconstruction methods can maximize the likelihood (or posterior) more directly, giving substantially lower reconstruction noise for noise levels where the B modes are signal dominated~\cite{Hirata:2003ka,Carron:2017mqf}. For measurement of the convergence power spectrum, gains on cosmological parameters from using iterative reconstruction are generally modest, because the well-resolved modes are dominated by cosmic variance of the lenses. For lensing rotation, where the signal is very small (and zero in the null-hypothesis of no rotation), the rotation reconstruction is noise dominated so the signal variance is negligible. Iterative reconstruction can therefore give larger improvements. For our purposes, we just need a forecast for the iterative lensing reconstruction noise, which can be calculated straightforwardly by iteration of the quadratic estimator formula using partially delensed fields: see Appendix~\ref{app:iters} for details.

\subsubsection{CMB experimental setups}
In this paper we consider different configurations of two near future CMB experiments, SO (at baseline and goal sensitivities), and CMB-S4. For modelling the effective CMB map level noise from all detectors, $N^{X}_l$, we use the harmonic-space Internal Linear Combination (ILC) model, described in \cite[]{SimonsObservatory:2018koc}, and include detector, beam, and foreground contaminants\footnote{For SO, the noise curves are taken from  \url{https://github.com/simonsobs/so_noise_models/tree/master/LAT_comp_sep_noise/v3.1.0}. These include baseline and goal sensitivities. For CMB-S4, the curves are taken from \url{http://sns.ias.edu/~jch/S4_190604d_2LAT_Tpol_default_noisecurves.tgz} and further details are available at \url{https://cmb-s4.uchicago.edu/wiki/index.php/Survey_Performance_Expectations}.}. We enforce effective cuts of $30\leq L \leq 5000$ for the polarization maps, and $30\leq L \leq 3000$ for the temperature. This is applied to all experimental configurations through setting $N^{ij}_l=\infty$ outside the specified range. The fractional sky coverage of both experiments is approximated as $f_{\textrm{sky}}=0.4$.

The noise on the GMV reconstructed observables are modelled at lowest order, i.e. with Eq.~\eqref{eq:N0} for $\hat{\omega}$, and a similar result for $\hat{\kappa}$ \cite[]{Maniyar:2021msb}. We also include forecasts with maximum a posteriori (iterative) estimators in our analyses, which have reconstruction noise calculated with the {\sc plancklens} code \cite[]{PL2018}. In general, we do not assume external tracers are used in the delensing step of the iterative reconstruction.

We also include a more speculative experimental setup, which we label `CMB-S4 deep'. Consider a CMB-S4 measurement covering a small patch of the sky, say $f_{\textrm{sky}}=0.05$. Long exposure to such a small area could reduce the polarization map level noise to $\Delta P=0.5\mu K$-arcmin. If we assume this particular patch is free of any significant foreground contamination then we can neglect foreground cleaning and simply model the noise by \cite[]{Knox:1995dq}
\begin{equation}
    N^{P}_L=\left(\frac{\Delta P\times10^{-6}}{T_{\mathrm{CMB}}}\right)^{2}\mathrm{exp}\left({\frac{L(L+1)\theta^2}{8\mathrm{ln}2}}\right),
    \end{equation}
which accounts for deconvolving a beam of full width half maximum (FWHM) $\theta = 2.3$ arcmin. It is harder to justify the exclusion of foreground modelling on temperature measurements, hence for this configuration we use polarization-only reconstruction. For consistency, we keep the multipole cuts of $30\leq L \leq 5000$.

\begin{figure}[t]
 	\includegraphics[width=\linewidth]{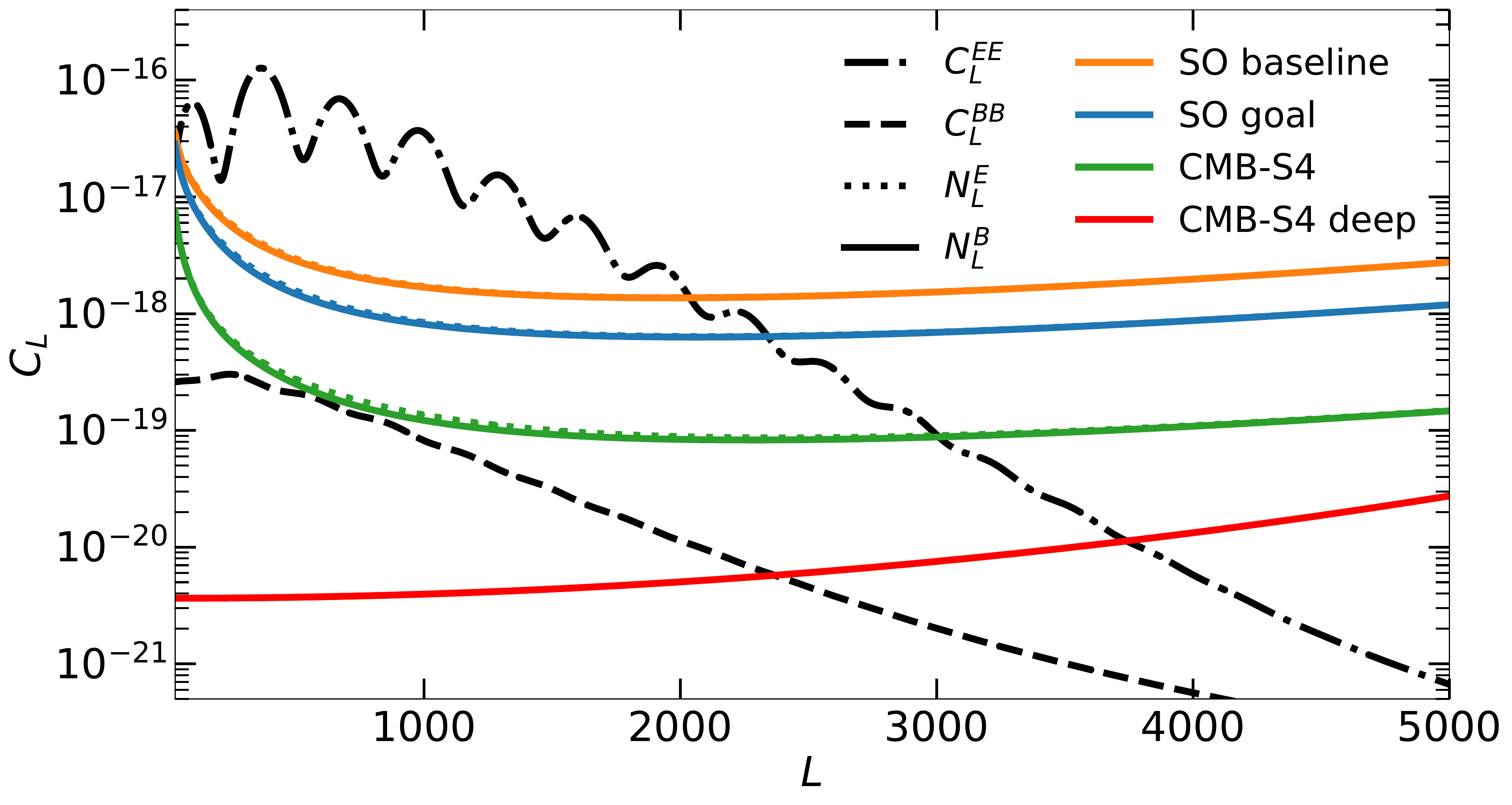}
	\caption{\small{The assumed effective CMB polarization noise power on E-mode (solid lines) and B-mode (dotted lines) measurements for different SO- and CMB-S4-like experimental setups. The lensed E-mode power-spectrum (dot-dashed line) is signal dominated on large scales for all configurations considered. However, the lensed B-mode spectrum (dashed line) is noise dominated on all scales with the wide area configurations.
CMB-S4 deep patch assumes noise levels significantly below the B-mode power on large scales, but only covers $5\%$ of the sky.
The deep patch assumes the limiting best case of white noise with no increase in noise due to foreground cleaning.
}}
	\label{fig:pol_noise}
\end{figure}

Fig.~\ref{fig:pol_noise} shows the polarization noise power spectra expected for the different experimental setups. The CMB-S4 deep can be considered a likely limiting best case scenario for the CMB-S4 lensing reconstruction noise. It is the only configuration with signal-dominated $C^{BB}_L$ at any scale, and is therefore ideal for iterative reconstruction methods. The large-patch foreground-cleaned configurations, by contrast, provide conservative results due to the substantial temperature and polarization foreground power assumed over the entire area.

\subsection{CMB rotation forecasts}\label{sec:set_ex}

It is instructive to consider the theoretical limit on the rotation signal-to-noise ratio, $S/N$, for a given experiment. We do this by constructing forecasts via the Fisher information \cite[]{Fisher:1922saa,Tegmark:1997rp}. Using curl reconstruction from observed lensed CMB maps, we assess detectability by how well we can measure the amplitude of the reconstructed rotation power spectrum, $C_L^{\hat{\omega}\hat{\omega}}$, compared to the fiducial model. Approximating the reconstruction noise as Gaussian and isotropic, and working in the flat-sky approximation, the Fisher estimate reduces to \cite[]{Hu:2000ee}

\begin{equation}\label{eq:forecast}
    \left(\frac{S}{N}\right)^2_{\hat{\omega}\hat{\omega}} \leq f_\textrm{sky}\int LdL\frac{\left(C_{L}^{\omega\omega}\right)^2}{\left(C_L^{\omega\omega} + N^{\omega}_L\right)^2}.
\end{equation}
Here we only include reconstruction noise at lowest order $N^{\omega}_L=N^{\omega}_0(L)$. Inclusion of the next leading noise term, $N^{\omega}_1$, could be important on certain scales, and degrade the forecasts further. However, it is still expected to be subdominant to $N^{\omega}_0$ and is not considered here. The integral is evaluated over the multipole range $30\leq L \leq 4000$.

The significance at which the reconstructed CMB lensing rotation power spectrum could be detected is shown in the first column of Table \ref{tab:fish_Cl} for the GMV estimator\footnote{Note, as the CMB-S4 deep configuration only uses polarization maps, the quadratic estimation is technically not a GMV reconstruction. However, the methods are identical, just without input from the temperature.}, and iterative reconstruction in brackets. It is clear the reconstructed rotation auto-spectrum will be very challenging to detect with the next generation of CMB experiments, in agreement with Ref.~\cite[]{Pratten:2016dsm}: the detection significance for SO at goal sensitivity is $\leq 0.1\sigma$, and there is only marginal improvement for CMB-S4 with predicted best $S/N$ at $0.4\sigma$ or $0.5\sigma$ with deep patch (Ref.~\cite[]{Pratten:2016dsm} forecast $0.7\sigma$ for S4, however their multipole cuts were different, and they use a simpler noise model without foreground cleaning). Iterative reconstruction does provide significant improvement for the CMB-S4 deep patch configuration, in which the auto-spectrum is forecast to be (un)detectable at $1.7\sigma$. Here the B-mode lensing signal benefits from the low polarization noise, giving rise to efficient delensing during the iterative process.

\begin{figure}[t]
 	\includegraphics[width=\linewidth]{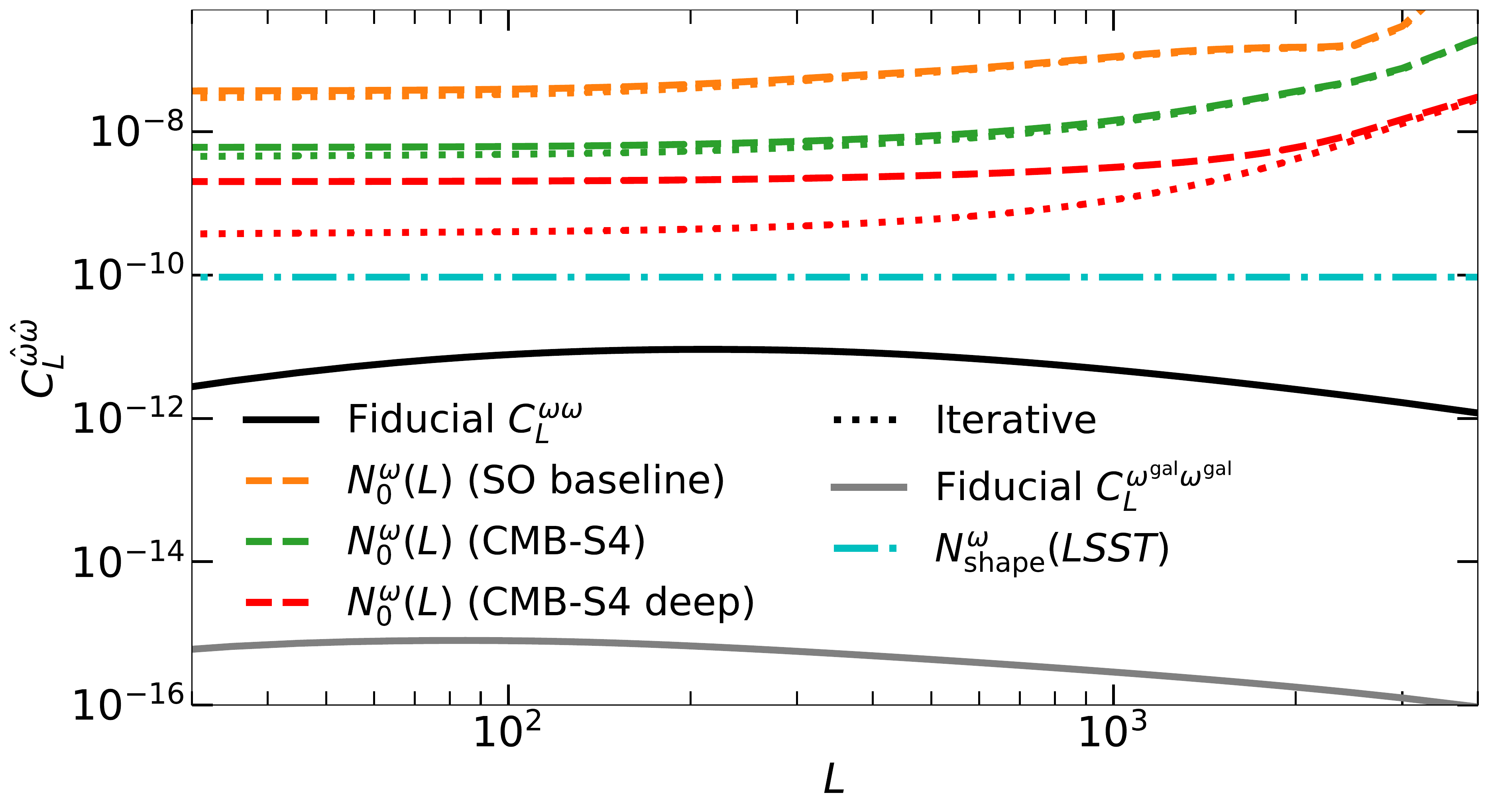}
	\caption{\small{The lowest-order reconstruction noise power, $N^{\omega}_0$, for GMV curl reconstruction (dashed lines), and iterative reconstruction (dotted lines) with SO- and CMB-S4-like experimental configurations. The CMB-S4 deep setup assumes polarization-only reconstruction, and has 5$\%$ sky coverage. The noise levels are typically $\sim 3$--$4$ orders of magnitude above the fiducial post-Born rotation power spectrum signal expected (solid black). The weak lensing shape noise for an LSST-like survey (red dot dashed line) is at least 5 orders of magnitude greater than the expected post-born rotation signal from galaxy lensing (solid grey).}}
	\label{fig:omega_ps}
\end{figure}

The difficulty of direct rotation detection is illustrated in Fig.~\ref{fig:omega_ps}: the reconstruction noise from the quadratic estimators is several orders of magnitude larger than the post-Born rotation power spectrum for all future CMB experiments considered.

\begin{table}[]
\resizebox{\linewidth}{!}{
\begin{tabular}{|l||c|c||c|c|c|}
    \hline
Experiment & $C^{\hat{\omega}\hat{\omega}}$ [$\sigma$] & $C^{\hat{\omega}\omega}$ [$\sigma$] & $C^{\hat{\omega}^{\textrm{gal}}\hat{\omega}^{\textrm{gal}}}$ [$\sigma$]& $C^{\hat{\omega}^{\textrm{gal}}\omega^{\textrm{gal}}}$ [$\sigma$]\\ \hline \hline
SO baseline & 0.0 (0.0) & 9.6 (9.8) & - & - \\
SO goal & 0.1 (0.1) & 11.7 (12.0) & - & - \\
CMB-S4 & 0.4 (0.4) & 23.5 (24.7) & - & - \\
CMB-S4 deep & 0.5 (1.7) & 17.9 (28.4) & - & - \\
LSST & - & - & 0.0 & 3.3 \\
Far future & - & - & 0.0 & 8.3 \\\hline
\end{tabular}}
\caption{\small{{Fisher forecasts for the expected detection significance (signal-to-noise ratios, $S/N$) of different field rotation power spectra with near future experiments/surveys. The first column is for the auto-spectrum of the CMB reconstructed field rotation, $\hat{\omega}$, and the second column shows the potential improvement achievable by crossing it with a perfect tracer for the rotation field. The $S/N$ numbers in brackets are the result of using iterative estimators instead of the GMV quadratic estimator. The CMB-S4 deep forecasts assume polarization-only reconstruction, and a fractional sky coverage of $f_\textrm{sky}=0.05$. The last two columns show the corresponding forecasts for observations of galaxy lensing field rotation, $\hat{\omega}^{\textrm{gal}}$, assuming a single broad redshift bin.}}\label{tab:fish_Cl}}
\end{table}

Now consider the cross correlation of the lensing reconstructed rotation, $\hat{\omega}$, with some external rotation template, $\hat{\omega}^{\mathrm{ext}}$. In the limit that this external rotation is a perfect template of the truth, i.e. $\hat{\omega}^{\mathrm{ext}}=\omega$, then the best case $(S/N)^2$ is now
\begin{equation}\label{eq:perfect_forecast}
    \left(\frac{S}{N}\right)^2_{\hat{\omega}\omega} \leq2f_\textrm{sky}\int LdL\frac{C_{L}^{\omega\omega}}{2C_L^{\omega\omega} + N^{\omega}_L}.
\end{equation}
The results in Table \ref{tab:fish_Cl} show this cross-correlation signal to be easily detectable with a sufficiently accurate external template for field rotation. In the limit that a perfect rotation template can be constructed, a significant detection of $12\sigma$ is possible for SO, and improves by a factor of $2$ for CMB-S4. Iterative reconstruction is of marginal importance at SO noise levels, however can improve the detectability by $>1\sigma$ for CMB-S4. Again, the deep patch CMB-S4 benefits significantly from iterative reconstruction, and can produce a cross-spectrum signal at $28.4\sigma$, $\sim 4\sigma$ greater than the wide CMB-S4 forecast.

The challenge to the detection of post-Born rotation on the CMB is now in the construction of a sufficiently accurate rotation template. The creation of such a template using density tracers is explored in Section~\ref{sec:template}.

\subsection{Galaxy Lensing Rotation}\label{sec:gal_omega}
Post-born lensing also contributes a non-vanishing curl component to galaxy lensing maps. However, galaxy lensing is mainly restricted to $z\lesssim 3$ as source galaxy numbers are only significant at low redshifts. It is therefore expected that post-Born corrections will be less important for galaxy lensing due to the reduced path length, and the rotation signal specifically will be weakened~\cite{Krause:2009yr}. Here we make a brief argument for focussing on lensing rotation with the CMB over galaxy lensing.

Assuming a standard galaxy shear estimator is used to measure shear B modes (equivalent to rotation), the main source of variance is from shape noise \cite[]{Hu:1999ek}
\begin{equation}\label{eq:shape_noise}
    N^{\omega^{\textrm{gal}}}_L \sim N^{s}_{\mathrm{shape}}=\frac{\sigma_{\epsilon}}{\bar{n}},
\end{equation}
where $\sigma_{\epsilon}=0.21$ is the standard deviation of galaxy ellipticities per component \cite[]{Maraio:2022ywi}, and $\bar{n}$ is the mean galaxy density per steradian. We consider two future galaxy surveys: the LSST gold sample with $\bar{n}=40$ galaxies per arcminute \cite[]{LSSTScience:2009jmu} and $f_{\textrm{sky}}=0.4$, and some ``far future'' full sky survey with $\bar{n}=100$ galaxies per arcminute. Both models share the same radial galaxy distribution described in Appendix \ref{app:tracers} by Eq.~\eqref{eq:gal_distro1} and Eq.~\eqref{eq:gal_distro2}. Then Eq.~\eqref{eq:forecast} and Eq.~\eqref{eq:perfect_forecast} produce the galaxy lensing rotation forecasts shown in the last two columns of Table \ref{tab:fish_Cl}.

As expected, the detection of the rotation auto-spectrum from galaxy surveys is even more challenging than for CMB experiments. For a near future LSST-like survey, a statistically significant measurement will be difficult even with a perfect rotation tracer. Realistic improvements to detectability require substantial reductions to the lensing noise, as the signal is more than 5 orders of magnitude smaller than current noise levels, see Fig. \ref{fig:omega_ps}. Near full sky coverage will also boost the signal as demonstrated with the ``far future" configuration, however even that is only forecast to have an upper limit of $8.3\sigma$ with a perfect external rotation tracer.

These simplistic forecasts only model the lensing signal from one effective redshift bin. Improvements could be gained with more realistic modelling of the curl signal from the full tomography expected for upcoming surveys. Other methods using additional polarization measurements for each galaxy source where available, such as those described in Ref.~\cite{Thomas:2016xhb}, could also improve galaxy rotation constraints. We leave these considerations for future work. For the remainder of this paper we focus on detecting field rotation from CMB lensing only, where the signal is larger and a detection is more easily achievable.

\section{Optimal Rotation Template}\label{sec:template}

In Section~\ref{sec:set_ex}, we showed that while the field rotation auto-spectrum will be undetectable for near-future CMB experiments, the correlation between CMB lensing and a suitable rotation template is detectable.  Post-Born rotation is generated at lowest order by two lensing events (see Section~\ref{sec:omega}). That is, image rotation is produced by the combination of a linear lensing shear followed by a subsequent (non-aligned) lensing shear at a lower redshift. The linear shear field is directly related to the convergence field, since they are both obtained as derivatives of an underlying scalar lensing potential, and the local source for the convergence depends on the matter density. This suggests that by combining tracers of the density that probe at least two different redshifts, we should be able to construct a template for the expected rotation signal. The CMB lensing convergence has significant contributions at all redshifts, so two copies of the convergence reconstruction could work.  External density tracers are generally  limited to lower redshifts ($z\alt 3$), but
can have much higher signal-to-noise and better redshift resolution. To extract the most information, we allow for general combinations of the convergence and external tracers.

\subsection{Template Construction}

Since the post-Born rotation template is expected to be quadratic in the density tracers, we can use a quadratic estimator in the form of Eq.~\eqref{eq:omega_qe}.
For a set of tracers $\{ \hat{a}^i\}$, we define a general
weighted quadratic estimator
\begin{equation}
    \hat{\omega}^{\mathrm{tem}}(\boldsymbol{L})\equiv \frac{F_L^{-1}}{2}\int\frac{d^2\bold{L}_1}{(2\pi)^2}G_{ij}(\boldsymbol{L}_1, \boldsymbol{L})\hat{a}^i(\boldsymbol{L}_1)\hat{a}^j(\boldsymbol{L}-\boldsymbol{L_1}),
\end{equation}
for some weights $G_{ij}$.
The normalization condition is set such that the cross-correlation with the true $\omega$ returns the fiducial post-Born rotation spectrum
\begin{equation}\label{eq:rot_qe_norm}
	\langle \hat{\omega}^{\textrm{tem}}(\boldsymbol{l}) \omega(\boldsymbol{l}')\rangle\equiv(2\pi)^2\delta_{D}(\boldsymbol{l}+\boldsymbol{l}')C^{\omega\omega}_l.
\end{equation}
Using this condition, and minimizing the variance to find the optimal weight functions, gives a normalization
\begin{equation}\label{eq:template_norm} F_L\!=\!\!\frac{1}{2C^{\omega\omega}_{L}}\!\!\int\frac{d^2\boldsymbol{L}_1}{(2\pi)^2}b^{\omega ij}_{(-\boldsymbol{L})\boldsymbol{L}_1\boldsymbol{L}_2}(\boldsymbol{C}_{\textrm{LSS}}^{-1})^{ip}_{L_1}(\boldsymbol{C}^{-1}_{\textrm{LSS}})^{jq}_{L_2}b^{\omega pq}_{(-\boldsymbol{L})\boldsymbol{L}_1\boldsymbol{L}_2},
\end{equation}
with the minimum variance template estimator given explicitly by
\begin{equation}\label{eq:flat_omega_QE_explicit}
    \hat{\omega}^{\textrm{tem}}(\bold{L})=\frac{F_L^{-1}}{2}\int\frac{d^2\bold{L}_1}{(2\pi)^2}b^{\omega ij}_{(-\boldsymbol{L})\boldsymbol{L}_1\boldsymbol{L}_2}\hat{\bar{a}}^i(\boldsymbol{L}_1)\hat{\bar{a}}^j(\boldsymbol{L}_2).
\end{equation}
Here we defined inverse covariance weighted fields $\bar{\boldsymbol{a}}\equiv(\boldsymbol{C}^{-1}_{\textrm{LSS}})\boldsymbol{a}$, with the covariance having signal and noise contributions $(\boldsymbol{C}_{\textrm{LSS}})^{ij}_L=C^{ij}_L+N^{ij}_L$.
The optimal weights are given in terms of the flat-sky rotation bispectra \cite[]{Lewis:2011au} defined by
\begin{equation}\label{eq:bi_def}
	\left\langle \omega(\boldsymbol{l})a_i(\boldsymbol{l}_1)a_j(\boldsymbol{l}_2)\right\rangle=(2\pi)^2\delta(\boldsymbol{l}+\boldsymbol{l}_1+\boldsymbol{l}_2)b^{\omega i j}_{\boldsymbol{l}\boldsymbol{l}_1\boldsymbol{l}_2}.
\end{equation}
As these bispectra have odd parity \cite[]{Pratten:2016dsm}, the moduli of the modes are insufficient to determine the bispectrum sign. Hence, the sign of the modes in Eq.~\eqref{eq:template_norm} and Eq.~\eqref{eq:flat_omega_QE_explicit} are important to determine the handedness of the triangle and ensure $\boldsymbol{L}_2=\boldsymbol{L}-\boldsymbol{L}_1$.

A useful consequence of the chosen normalization for the estimator in Eq.~\eqref{eq:rot_qe_norm} is that the variance of the estimator is
\begin{equation}
C^{\hat{\omega}^{\textrm{tem}}\hat{\omega}^{\textrm{tem}}}_L=C^{\omega\omega}_L+N^{\hat{\omega}^{\textrm{tem}}}_L=C^{\omega\omega}_LF_L^{-1}.
\end{equation}
 This in turn provides a relation between the normalization, $F_L$, and the correlation coefficient of the estimator with the true rotation
\begin{equation}\label{eq:rho_F_L}
	\rho_{\omega\hat{\omega}^{\textrm{tem}}}^2(L)=F_L.
\end{equation}

\subsection{Tracers}\label{sec:tracers}
We assume that density tracers used as input to the template estimator are Gaussian fields with uncorrelated noise. If we restrict the form of these tracers to
\begin{equation}\label{eq:density_tracer}
	a(\boldsymbol{L})=\int d\chi W_a(\chi)\delta(\boldsymbol{L},\chi),
\end{equation}
then they are fully described by their window function, $W_a$, which describe the radial sensitivity to the underlying matter distribution. The general form of the density auto- and cross-spectra at lowest order Limber (for $a\neq \kappa$) is then
\begin{equation}\label{eq:density_tracer_ps}
	C^{a_1a_2}_{L}=\int d\chi' \frac{W_{a_1}(\chi')W_{a_2}(\chi')}{(\chi')^2}P_{\delta\delta}\left(k\approx\frac{L}{\chi'},z(\chi')\right),
\end{equation}
or when the cross-spectrum involves one $\kappa$
\begin{equation}\label{eq:kappa_tracer_ps}
	C^{a\kappa}_{L}(\chi)=\int d\chi' \frac{W_{a}(\chi')W_{\kappa}(\chi', \chi)}{(\chi')^2}P_{\delta\delta}\left(k\approx\frac{L}{\chi'},z(\chi')\right).
\end{equation}
For completeness, the CMB convergence auto-spectrum is given in Eq.~\eqref{eq:kappa_ps}.

The rotation-density-density bispectra in Eq.~\eqref{eq:flat_omega_QE_explicit} can be written explicitly as
\begin{multline}\label{eq:rot_bi}
    \begin{split}
    b^{\omega ij}_{(-\boldsymbol{L})\boldsymbol{L}_1\boldsymbol{L}_2}=-2\frac{(\boldsymbol{L}_1\cdot \boldsymbol{L}_2)[\boldsymbol{L}_1
    \times \boldsymbol{L}_2]}{(L_1)^2(L_2)^2}\big[\mathcal{M}^{ij}(L_1,L_2)\\-\mathcal{M}^{ji}(L_2,L_1) \big],
        \end{split}
\end{multline}
with $\boldsymbol{L}_2=\boldsymbol{L}-\boldsymbol{L}_1$, where the mode couplings are defined as
\begin{multline}\label{eq:mode_coupling}
    \begin{split}
    \mathcal{M}^{a_1a_2}(L,L')\equiv\int^{\chi_*}_{0}d\chi \frac{W_{\kappa}(\chi,\chi_*)W_{a_1}(\chi)}{\chi^2}\\
    P_{\delta\delta}\left(k\approx\frac{L}{\chi},z(\chi)\right)C^{a_2\kappa}_{L'}(\chi).
        \end{split}
\end{multline}
As we are now only considering CMB lensing, it is implicitly assumed that $\chi_s=\chi_*$. Also, in the case of $a_2=\kappa$ in Eq.~\eqref{eq:mode_coupling}, $C^{a_2\kappa}_{L'}(\chi)=C^{\kappa\kappa}_{L'}(\chi, \chi_*)$ which is given in Eq.~\eqref{eq:kappa_ps}. The form of these bispectra are in general agreement to the similarly-defined post-Born convergence bispectra with two external tracers \cite[]{Fabbian:2019tik}.

These rotation bispectra should be individually more detectable than the rotation reconstruction auto-spectrum, as good density tracers will not suffer from the severe noise domination that afflicts the lensing-reconstructed $\hat{\omega}$. This has already been demonstrated for the rotation-convergence-convergence bispectrum in Ref.~\cite[]{Pratten:2016dsm}.

\subsubsection{Tracer choices}\label{sec:tracer_discussion}
\begin{figure}[t]
 	\includegraphics[width=\linewidth]{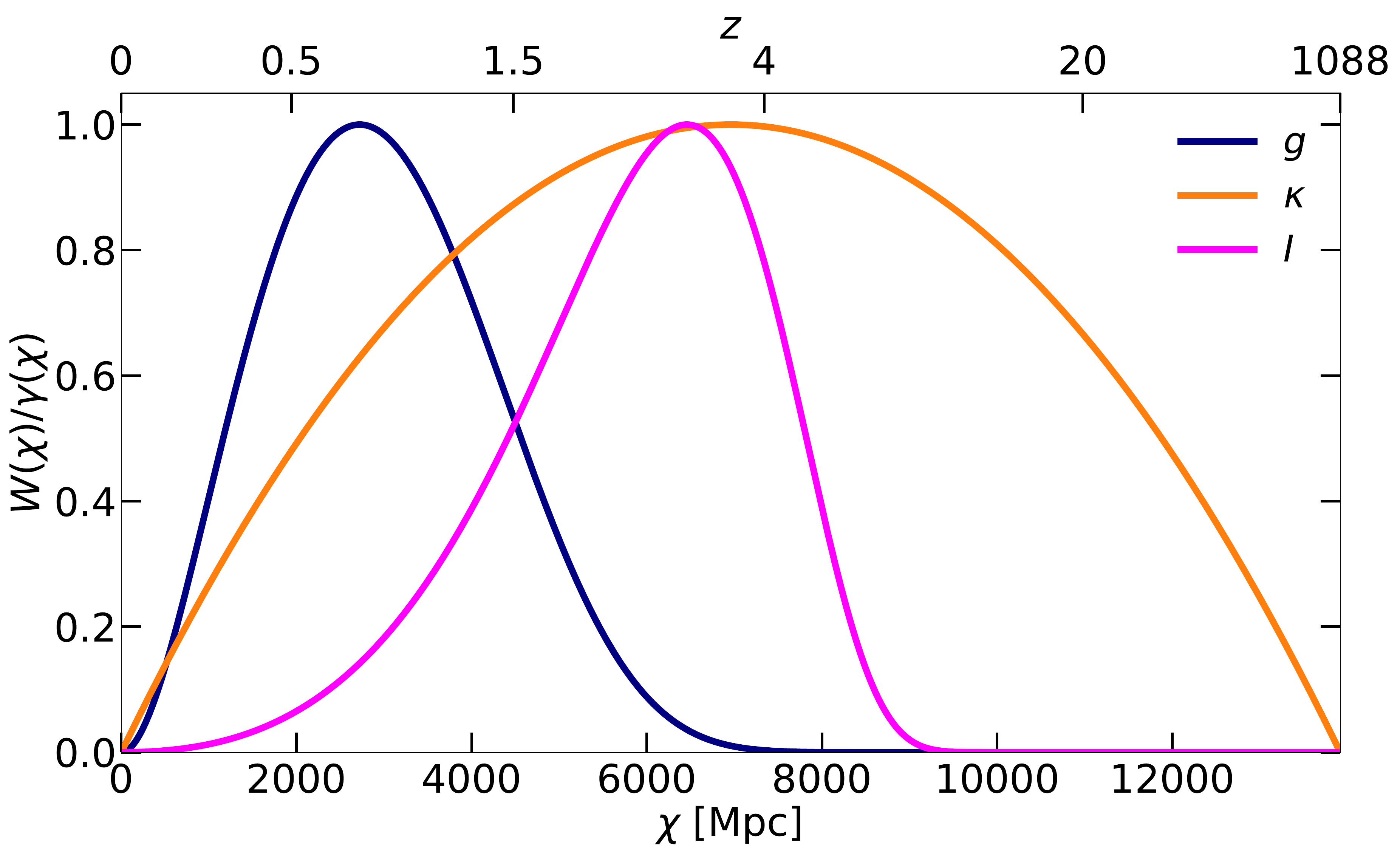}
	\caption{\small{The Window functions, $W$, for tracers of the underlying matter distribution: CMB lensing convergence ($\kappa$), galaxy counts ($g$), and CIB ($I$). All are functions of comoving radial distance, $\chi$, with maxima normalized to unity, and have been converted into kernels for the Weyl gravitational potential via Poisson factor, $\gamma$, defined in Eq.~\eqref{eq:poisson}. In matter domination the Weyl potential is constant (before decaying slightly once dark energy is important), so the figure approximately illustrates the relative contributions of the different redshift regions traced by each observable.
}}
	\label{fig:windows}
\end{figure}

We consider three density tracers as input for the rotation template: the CMB lensing convergence, $\kappa$, galaxy clustering, $g$, and the CIB, $I$. The modelling of these tracers, specifically of their window functions, along with their noise contributions are detailed in Appendix \ref{app:tracers}. Combinations of tracers that probe different redshift regions are ideal for a post-Born rotation template, as they are better able to mimic the lens-lens coupling. From the window functions plotted in Fig.~\ref{fig:windows}, it is clear the galaxy clustering observable probes the low redshift universe, only sensitive to the underlying matter field for $z\lesssim3$. Here we consider unbinned $g$, however tomographic binning of the galaxy counts would in principle provide a better probe of the contributions to the rotation signal from low redshifts. The CIB has sensitivity peaked at higher redshift. As the last scattering surface is in the early universe, CMB fields are lensed by density perturbations over almost the entire redshift range, as seen in the broad window function. The lensing kernels determining the lens-lens coupling are broad in redshift, so combinations of these tracers should provide reasonable coverage of the pairs of redshifts that contribute to the rotation signal. We do not consider cosmic shear as one of the input tracers here, though in principle it could also be used as a lower-redshift tracer.

\begin{figure}[t]
 	\includegraphics[width=\linewidth]{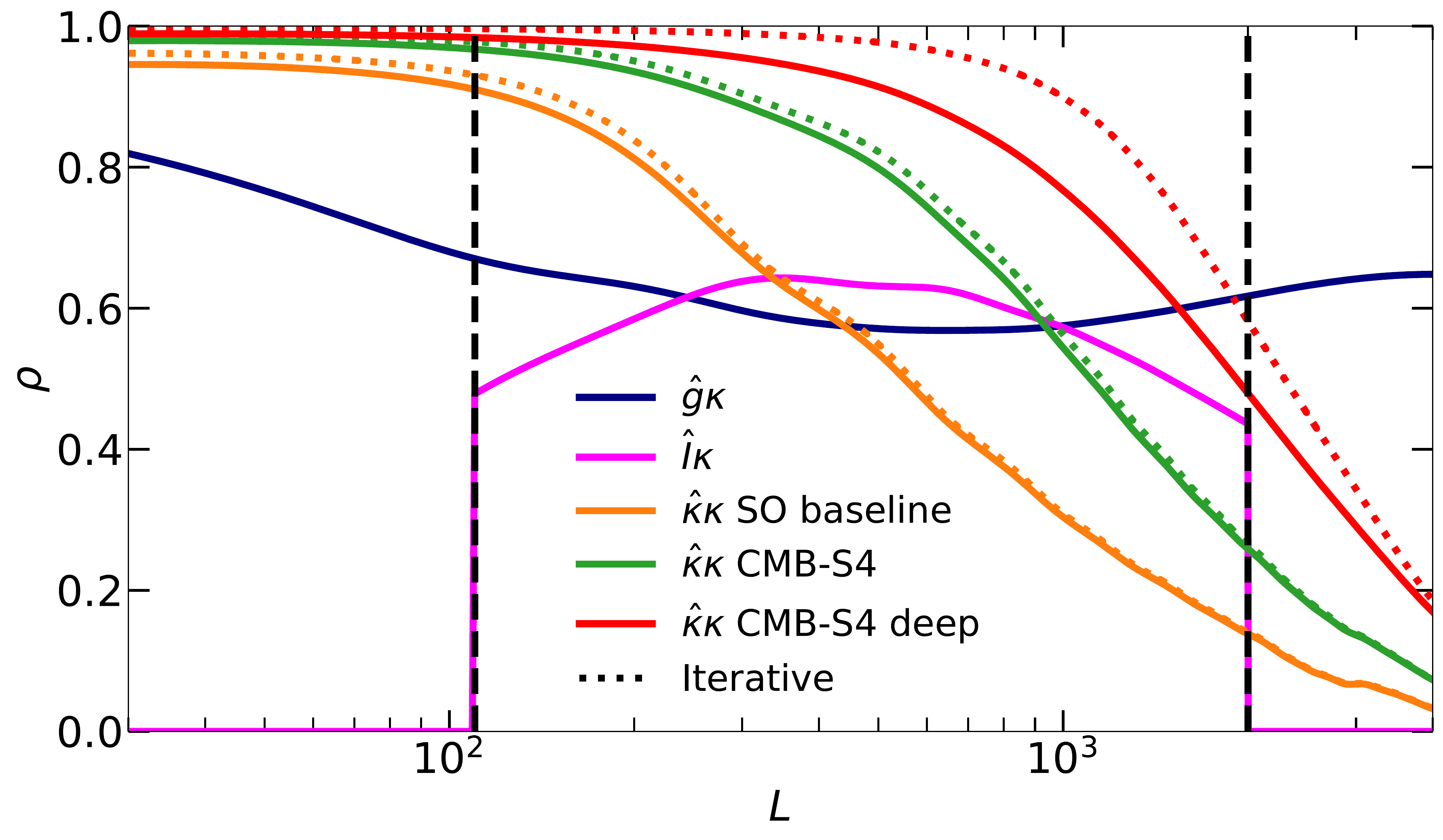}
	\caption{\small{The correlation coefficient, $\rho$, of the fiducial CMB lensing convergence, $\kappa$, with LSS density observables: LSST gold sample galaxy counts, $\hat{g}$, and Planck $353$ GHz CIB, $\hat{I}$. The CIB observable is shown with multipole cuts at $110\leq L\leq2000$, illustrated with the dashed lines. The coefficient between $\kappa$ and the GMV reconstructed CMB lensing convergence, $\hat{\kappa}$, is shown for SO at baseline sensitivity and CMB-S4. The improved correlation obtained using iteratively-reconstructed $\hat{\kappa}$ is shown in the dotted lines. The CMB-S4 deep configuration assumes polarization-only reconstruction.}}
	\label{fig:corr}
\end{figure}

For $\hat{\kappa}$, we continue to use the expected experimental configurations for SO and CMB-S4. The LSST gold model \cite{LSSTScience:2009jmu} is the natural choice for $\hat{g}$, as the survey sky coverage will optimally overlap with SO and both surveys are expected to be operational at around the same time period. Finally, $\hat{I}$ is modelled from the currently-available Planck CIB $353$ GHz map. Fig. \ref{fig:corr} shows the correlation coefficients of the different tracers with the true CMB lensing convergence, given the assumed experimental configuration
\begin{equation}\label{eq:general_rho}
	\rho_{\hat{a}\kappa}(L)=\frac{C^{a\kappa}_L}{\sqrt{(C^{aa}_L + N^a_L)C^{\kappa\kappa}_L}}.
\end{equation}
The SO $\hat{\kappa}$ becomes noise dominated at high L, resulting in the poor correlation on small scales. This is improved by CMB-S4, however the trend is still visible. It is also seen that iterative reconstruction is most important for CMB-S4 deep patch, due to the expected dominance of the B-mode lensing signal. Galaxy clustering is signal dominated on all scales due to the high expected LSST galaxy number density, reflected in the consistently strong correlation with $\kappa$. The CIB has better correlation on small scales than $\hat{\kappa}$ from the wide patch CMB experiments, however we apply strict modal cuts to the CIB of $110 \leq L\leq 2000$ due to uncertainty in modelling the dust and instrumental noise, see Appendix \ref{app:tracers} for more details. The correlation coefficients are consistent with Refs.~\cite[]{Yu:2017djs,Namikawa:2021gyh}.

\subsection{Fast Real-space Template Construction}
The rotation template of Eq.~\eqref{eq:flat_omega_QE_explicit} involves a computationally expensive double integral over a pair of weighted external density maps. The estimator's response functions are rotation bispectra, $b^{\omega ij}$, which from Eq.~\eqref{eq:rot_bi} and Eq.~\eqref{eq:mode_coupling} involve radial integration over terms separable in $\boldsymbol{L}_1$ and $\boldsymbol{L}_2$. Therefore, at each step along this integral, the template actually takes the form of a (weighted and normalized) Fourier-space convolution. We exploit this by using convolution theorem to write the integrand of the template as configuration-space products of mode-separated functions. Denoting a Fourier transform with, $\mathcal{F}$, the template can be quickly computed in the form

\begin{equation}\label{eq:omega_qe_split_ft}
\begin{aligned}
\hat{\omega}^{\textrm{tem}}(\boldsymbol{L})=-2F_L^{-1}\int^{\chi_*}_0d\chi \frac{W_{\kappa}(\chi, \chi_*)}{\chi^2}\delta_{pq}\epsilon_{rs}\mathcal{F}\left[ \Gamma^{pr}\Lambda^{qs}\right](\boldsymbol{n},\chi),
\end{aligned}
\end{equation}
where the functions $\Gamma = \mathcal{F}^{-1}(\gamma)$, and $\Lambda = \mathcal{F}^{-1}(\lambda)$ are defined as
\begin{equation}
\gamma^{pr}(\boldsymbol{L},\chi)\equiv\frac{L^pL^r}{L^2}W_i(\chi)P_{\delta\delta}\left(k\approx\frac{L}{\chi},z(\chi)\right)(\boldsymbol{C}^{-1}_{\textrm{LSS}})^{ij}_La_j({\boldsymbol{L}}),
\end{equation}
\begin{equation}
\lambda^{pr}(\boldsymbol{L},\chi)\equiv\frac{L^pL^r}{L^2}C^{i\kappa}_L(\chi)(\boldsymbol{C}^{-1}_{\textrm{LSS}})^{ij}_La_j({\boldsymbol{L}}).
\end{equation}
Here $L^i$ is component $i$ of the vector $\boldsymbol{L}$. The template is now computationally tractable. Convergence is achievable with naive uniform sampling of $\sim100$ steps in $\chi$, which can be easily reduced with targeted sampling along the radial path.

\section{Forecasts}\label{sec:fisher}

\subsection{Bispectra Forecasts}
Our aim is to probe the CMB post-Born rotation power spectrum, which we estimate through the cross-correlation of the CMB lensing reconstructed rotation, $\hat{\omega}$, with an externally constructed template, $\hat{\omega}^{\mathrm{tem}}$. From Eq.~\eqref{eq:flat_omega_QE_explicit},  it is apparent that $C^{\hat{\omega}\hat{\omega}^{\textrm{tem}}}_L$ is dependent on weighted combinations of observed bispectra $b^{\hat{\omega}\hat{\imath}\hat{\jmath}}_{\boldsymbol{L}\boldsymbol{L}_1\boldsymbol{L}_2}$. It is therefore useful to understand the relative contributions that each rotation bispectrum provides toward the rotation power spectrum detection significance.

We consider forecasts of the $S/N$ for the individual rotation bispectra via the flat sky Fisher information, where the parameter of interest is the amplitude of the bispectrum
\begin{equation}\label{eq:fish_bi}
\begin{aligned}
&\left(\frac{S}{N}\right)^2_{\hat{\omega} \hat{a}_1\hat{a}_2} \leq\frac{f_\textrm{sky}}{\pi}\int\frac{d^2\boldsymbol{L}_1d^2\boldsymbol{L}_2}{(2\pi)^2}\\
    &\frac{(b^{\omega a_1a_2}_{\boldsymbol{L}\boldsymbol{L}_1\boldsymbol{L}_2})^2}{\left[(\boldsymbol{C})^{a_1a_1}_{L_1}(\boldsymbol{C})^{a_2a_2}_{L_2}+(\boldsymbol{C})^{a_1a_2}_{L_1}(\boldsymbol{C})^{a_1a_2}_{L_2}\right](C^{\omega\omega}_{L}+N^{\omega}_L)}.
\end{aligned}
\end{equation}
As for the power spectra forecasts, the covariance matrix elements are $(\boldsymbol{C})_L^{a_1a_2}=(\boldsymbol{C}_{\textrm{LSS}})_L^{a_1a_2}=C_L^{a_1a_2}+N_L^{a_1a_2}$, and the integrals are evaluated over multipole range $30\leq L\leq4000$, with noise taken to be uncorrelated between observables, i.e. $N^{a_1a_2}_L=0$ if $a_1\neq a_2$.

\begin{table}[]
\resizebox{\linewidth}{!}{
\begin{tabular}{|l||l|l|l|l|l|l|}
    \hline
Experiment & $b^{\hat{\omega}\hat{\kappa}\hat{\kappa}}$ [$\sigma$] & $b^{\hat{\omega} \hat{g}\hat{g}}$ [$\sigma$]& $b^{\hat{\omega} \hat{g}\hat{\kappa}}$[$\sigma$] & $b^{\hat{\omega} \hat{I}\hat{I}}$ [$\sigma$]& $b^{\hat{\omega}\hat{I}\hat{\kappa}}$ [$\sigma$]& $b^{\hat{\omega} \hat{I}\hat{g}}$ [$\sigma$]\\ \hline \hline
SO baseline & 0.4 (0.5) & 0.2 (0.3) & 2.8 (2.9) & 0.1 (0.1) & 0.4 (0.5) & 3.5 (3.5)\\
SO goal & 0.6 (0.7) & 0.2 (0.3) & 3.8 (3.9) & 0.1 (0.1) & 0.6 (0.6) & 4.3 (4.4)\\
CMB-S4  & 1.8 (2.0) & 0.6 (0.6) & 9.5 (10.3) & 0.2 (0.2) & 1.4 (1.6) & 8.8 (9.3)\\
CMB-S4 deep & 1.5 (3.2) & 0.5 (0.7) & 8.2 (14.8) & 0.2 (0.3) & 1.1 (2.1) & 6.5 (11.0)\\ \hline
\end{tabular}}
\caption{\small{Fisher forecasts for the detection significance (signal-to-noise ratios, $S/N$) of various mixed lensing rotation-tracer-tracer bispectra individually. The CMB field rotation, $\hat{\omega}$, and convergence, $\hat{\kappa}$, are reconstructed with the GMV estimator (or iterative method in brackets). The forecasts are given for noise levels expected of SO (at baseline and goal sensitivities) and CMB-S4. The additional CMB-S4 deep patch experiment uses polarization only reconstruction, and covers just 5$\%$ of the sky. The other observables are LSST gold sample galaxy counts, $\hat{g}$, and the Planck CIB measurements, $\hat{I}$, at $353$ GHz.}}
\label{tab:fish_bi}
\end{table}

The resulting upper-bound $S/N$ for the different bispectra are given in Table \ref{tab:fish_bi}. The pure CMB rotation bispectrum $b^{\omega\kappa\kappa}$ has an expected $S/N$ of $3.2\sigma$ for iteratively reconstructed CMB-S4 deep patch, while the more conservative CMB-S4 wide result is only $2.0\sigma$. Therefore, it will remain challenging for near-future CMB measurements alone to detect $\omega$ with any significance due to large reconstruction noise levels.

\begin{figure}
 	\includegraphics[width=\linewidth]{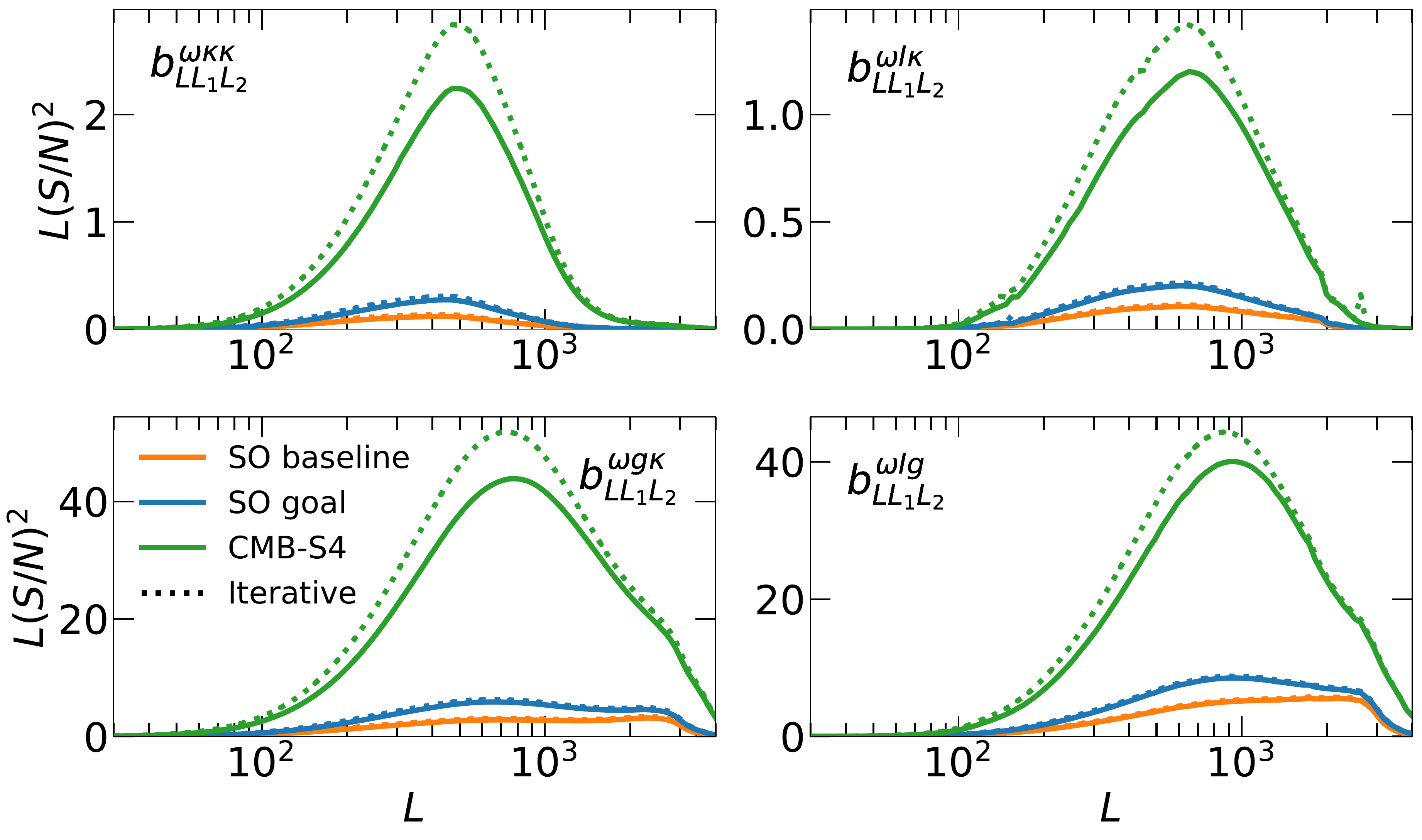}
	\caption{\small{Contributions to the Fisher information as a function of the $\omega(\vL)$ rotation multipole ($L$) for selected bispectra from Table \ref{tab:fish_bi}. The curves are scaled by $L$ to avoid contributions appearing misleading on smaller scales due to the choice of semi-logarithmic axes. The solid lines represent CMB reconstruction using GMV estimators, while the dotted show iterative reconstruction.}}
	\label{fig:bi_fish}
\end{figure}

The most detectable bispectra combine a low redshift tracer with a high redshift one. It follows that bispectra involving galaxy density combined with either the CIB or $\kappa$, are thereby easier to detect. For SO baseline sensitivity, $b^{\hat{\omega} \hat{g}\hat{\kappa}}$ and $b^{\hat{\omega} \hat{g}\hat{I}}$ have signals at $2.8\sigma$ and $3.5\sigma$ significance. This is increased by $\sim1\sigma$ if the goal sensitivity is reached. For CMB-S4, the detection significance improves significantly to $10.3\sigma$ and $9.3\sigma$ with iterative reconstruction. This is further improved with a deep patch to $14.8\sigma$ and $11.0\sigma$. Bispectra that combine observables occupying similar redshift space are unlikely to be detected even by CMB-S4. This is especially true for combinations of the same observable, but also for $b^{\hat{\omega} \hat{I}\hat{\kappa}}$ due to the significant overlap between the CIB and $\kappa$ in redshift sensitivity as is shown in Fig. \ref{fig:windows}.

The Fisher contributions per log mode for each individual bispectra are shown in Fig. \ref{fig:bi_fish} for SO and CMB-S4 noise levels. The peak of the contributions to $S/N$ for a pair of $\kappa$ tracers is centred at $L\sim500$, with very little contribution beyond $L>2000$ due to lensing reconstruction noise dominance on small scales, as illustrated in Fig. \ref{fig:kappa_ps}. In contrast, the expected LSST galaxy density is signal dominated to high L, and hence bispectra involving $g$ have significant contributions to $S/N$ at $L>1000$. For bispectra involving the CIB, contributions outside the stricter limits of $110\leq L\leq2000$ are suppressed.

\subsection{Cross-spectra Forecasts}

We can now finally quantify the ability of the post-Born rotation template, $\hat{\omega}^{\mathrm{tem}}$, to reproduce the underlying field rotation spectrum when crossed with the noise-dominated CMB lensed rotation estimator, $\hat{\omega}$. The flat-sky Fisher for this rotation cross spectrum is
\begin{equation}\label{eq:fish_opt}
    \left(\frac{S}{N}\right)^2_{\hat{\omega}\hat{\omega}^{\textrm{tem}}} \leq\frac{f_{\mathrm{sky}}}{\pi}\int d^2\bold{L}\frac{C^{\omega\omega}_L}{(C^{\omega\omega}_L + N^{\omega}_L)F_L^{-1} + C^{\omega\omega}_L}.
\end{equation}
In the limit of a perfectly reconstructed template, i.e. $F_L=1$, we recover the Fisher information for the rotation cross-spectrum with a perfect tracer (Eq.~\ref{eq:perfect_forecast}) as expected. Hence, the limits originally set in Section~\ref{sec:set_ex} hold true. However, if the template reconstruction is not perfect and the lensing reconstructed $\hat{\omega}$ is noise dominated, the Fisher information reduces to
\begin{equation}\label{eq:fisher_omega_opt}
    \left(\frac{S}{N}\right)^2_{\hat{\omega}\hat{\omega}^{\textrm{tem}}} \lesssim \frac{f_{\mathrm{sky}}}{\pi}\int d^2\bold{L}\frac{C^{\omega}_L F_L}{N^{\omega}_L}.
\end{equation}
This is the (flat sky equivalent) Fisher matrix of the standard optimal bispectrum estimator (see e.g. Ref.~\cite[]{Lewis:2011fk}).

\begin{figure}[t]
 	\includegraphics[width=\linewidth]{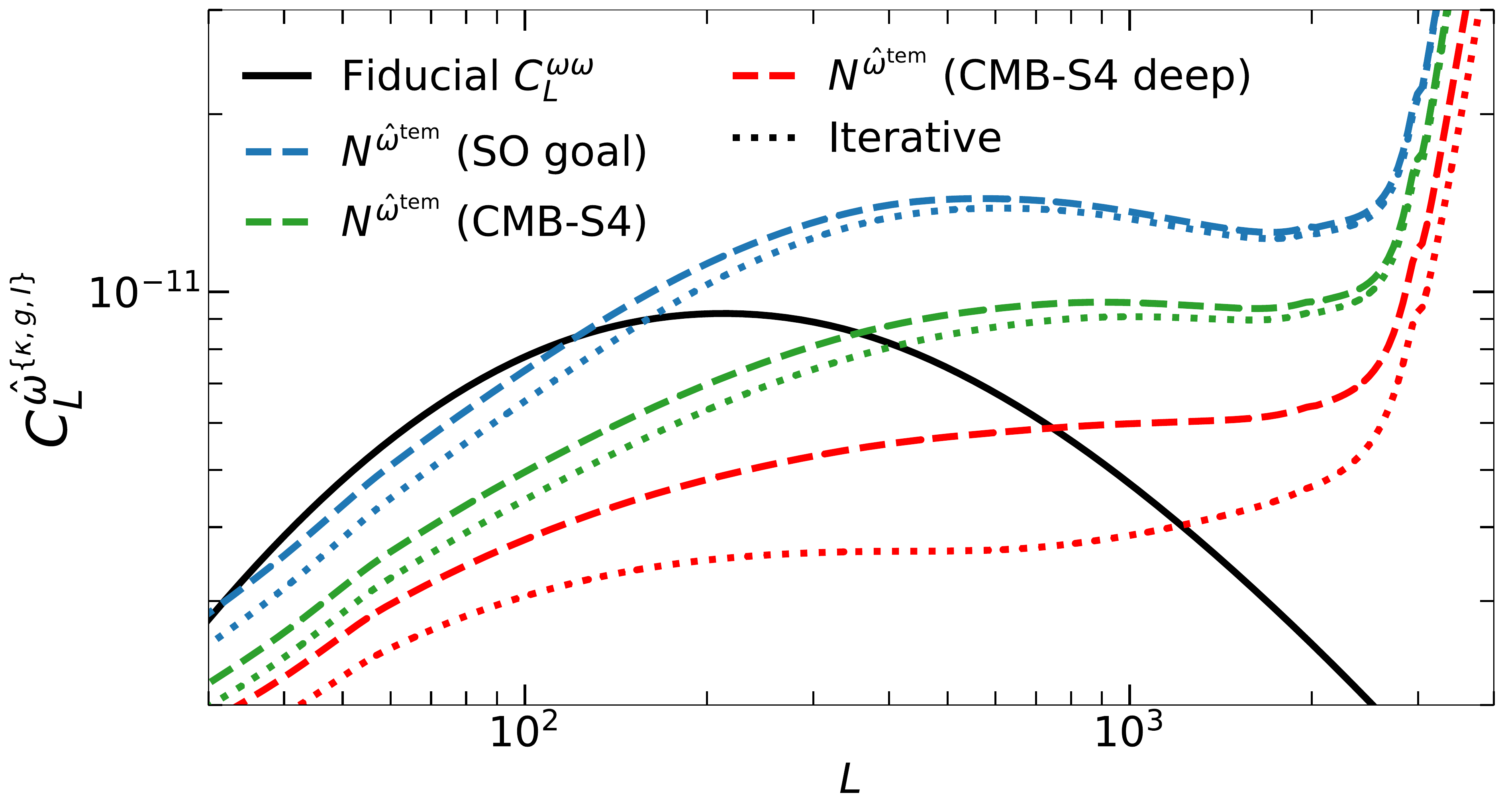}
	\caption{\small{Noise levels (dashed lines) on the auto-spectrum of a rotation template, $\hat{\omega}^{\textrm{tem}}$, for SO at goal sensitivity, CMB-S4, and CMB-S4 deep patch. Dotted lines show corresponding results using iterative reconstruction on $\hat{\kappa}$ instead of GMV. Polarization-only reconstruction is used for CMB-S4 deep patch lensing observables, over a sky fraction of just $5\%$. The template is constructed from the optimal combination of all tracers considered in this paper, $\hat{a}\in\{\hat{\kappa},\hat{g},\hat{I}\}$. On large scales the template noise is low compared to the expected fiducial post-Born field rotation signal (solid black), but the template is noise-dominated on small scales.}}
	\label{fig:omega_lss_ps}
\end{figure}

\begin{figure}[t]
 	\includegraphics[width=\linewidth]{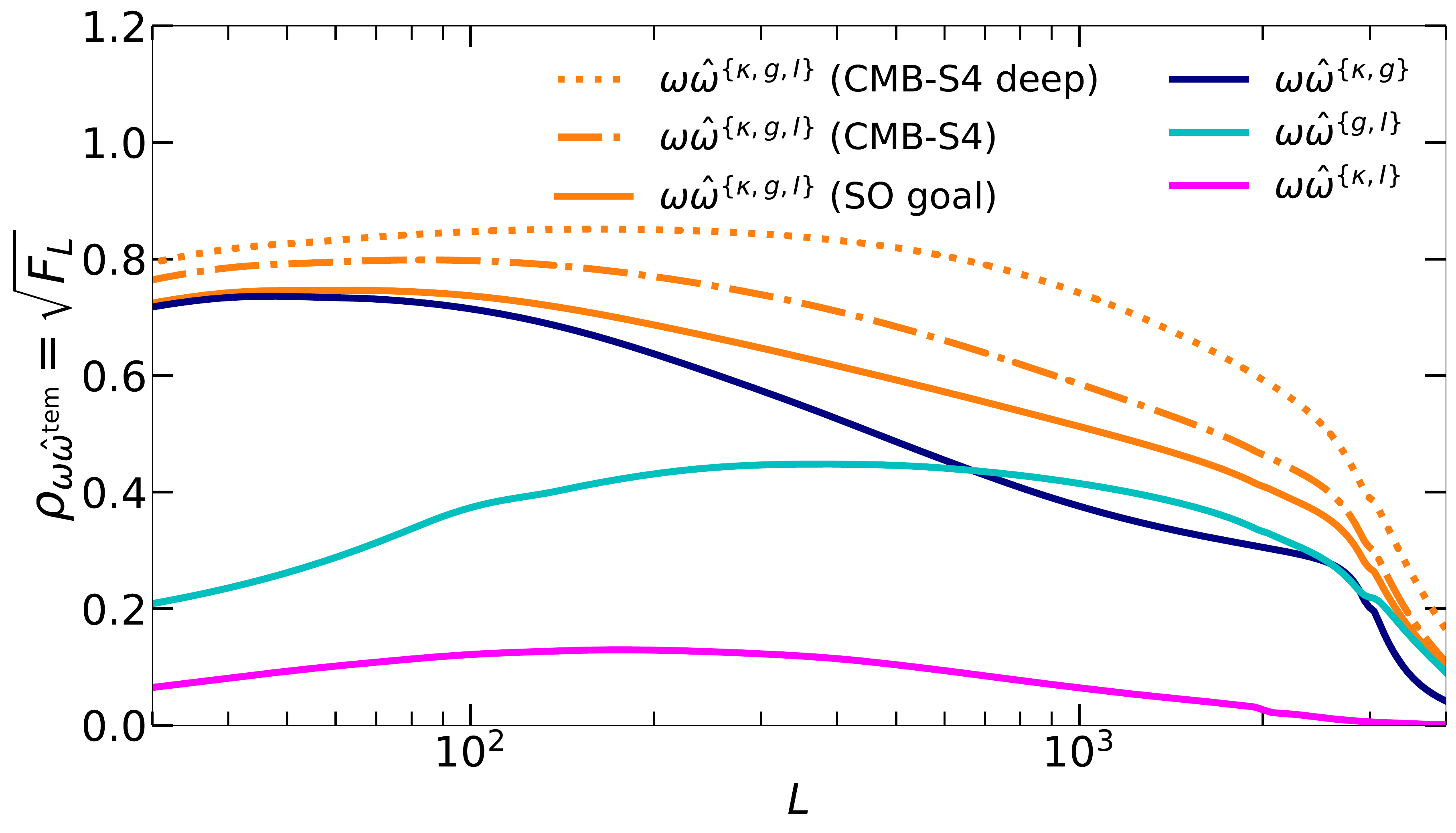}
	\caption{\small{The correlation coefficients, $\rho$, for different rotation templates with the true fiducial post-Born field rotation, calculated using Eq.~\eqref{eq:rho_F_L}.
All results assume iterative reconstruction for the lensing convergence tracer, either with noise levels for SO goal (solid lines), CMB-S4 (dash-dotted line), or CMB-S4 deep (dotted line). CMB-S4 deep uses only polarization in the lensing reconstructions.}}
	\label{fig:omega_corr}
\end{figure}

Does the template get close to a near-perfect reconstruction of the field rotation signal? Fig. \ref{fig:omega_lss_ps} plots the noise levels on the auto-spectrum of a template using all three density tracers, i.e. $\hat{a}\in\{\hat{\kappa}, \hat{g}, \hat{I}\}$, which we will denote $\hat{\omega}^{\{\kappa,g,I\}}$. It shows the noise levels are of similar magnitude to the estimated signal, a significant improvement compared to the noise-dominated CMB lensing reconstruction in Fig. \ref{fig:omega_ps}. For CMB-S4 deep, the noise only begins to dominate at $L>1000$ for a template using iteratively reconstructed $\hat{\kappa}$. However, noise does begin to dominate the template at lower $L$ for the large patch experiments. Fig. \ref{fig:omega_corr} shows the correlation coefficients of different templates with the fiducial rotation, $\rho_{\omega\hat{\omega}^{\textrm{tem}}}$, as defined in Eq.~\eqref{eq:rho_F_L}. As expected, the best correlations are achieved when all the density tracers are used. The correlations are greatest for $L\lesssim1000$, however $F_L$ does not exceed $0.73$ for any of the sensitivities considered.
For near future experiments, $\hat{\omega}^{\textrm{tem}}$ will not achieve a close-to-perfect emulation at any mode, but should still recover significant portions of the rotation signal.

As lensing reconstruction is completely noise dominated, $C_L^{\omega\omega}\ll N^{\omega}_L$, the Fishers used to forecast the upper-bound $S/N$ on the rotation cross-spectra, Eq.~\eqref{eq:fish_opt} and Eq.~\eqref{eq:fisher_omega_opt}, are equivalent. The results can therefore also be interpreted as forecasts for optimal rotation-density-density bispectra, and are shown in Table \ref{tab:fish_opt} for different sets of tracers. A statistically significant detection of $>5\sigma$ is now feasible with SO. Using LSST $\hat{g}$, the Planck $353$ GHz $\hat{I}$, and CMB reconstructed $\hat{\kappa}$, the forecast reconstruction of the rotation spectrum is $4.5\sigma$--$5.7\sigma$ for SO baseline--goal sensitivities. This is improved by $0.2\sigma$--$0.4\sigma$ with iterative lensing reconstruction. For the CMB-S4 wide configuration, the signal can be probed at $13.6\sigma$--$14.7\sigma$ depending on the lensing reconstruction method used. This is improved significantly with the deep patch configuration, resulting in the best forecast of $21.3\sigma$ when iterative methods are used for the lensing observables.

\begin{table}[]\resizebox{\linewidth}{!}{
\begin{tabular}{|l||l|l|l|l|}
    \hline
Experiment & $C^{\hat{\omega}\hat{\omega}^{\{\kappa,g\}}}$ [$\sigma$]& $C^{\hat{\omega}\hat{\omega}^{\{\kappa,I\}}}$ [$\sigma$]& $C^{\hat{\omega}\hat{\omega}^{\{g,I\}}}$ [$\sigma$]& $C^{\hat{\omega}\hat{\omega}^{\{\kappa,g,I\}}}$ [$\sigma$]\\ \hline \hline
SO baseline & 3.3 (3.5) & 0.6 (0.6) & 3.6 (3.7) & 4.5 (4.7)\\
SO goal & 4.5 (4.9) & 0.8 (0.9) & 4.4 (4.6) & 5.7 (6.1) \\
CMB-S4 & 12.1 (13.3) & 2.1 (2.4) & 9.3 (9.8) & 13.6 (14.7)\\
CMB-S4 deep & 10.7 (20.4) & 1.8 (3.7) & 6.9 (11.6) & 11.3 (21.3)\\ \hline
\end{tabular}}
\caption{\small{Fisher forecasts for the detection significance (signal-to-noise ratios, $S/N$) of the cross-spectrum between the CMB reconstructed field rotation, $\hat{\omega}$, and different optimally-constructed rotation templates, $\hat{\omega}^{\textrm{tem}}$. The density tracers used as input to the template are labelled in the superscript, e.g. the template denoted $\hat{\omega}^{\{\kappa,g,I\}}$ is optimally constructed from all combinations of the CMB lensing convergence, $\hat{\kappa}$, galaxy clustering, $\hat{g}$, and CIB, $\hat{I}$. The results are given for different CMB experimental setups assuming GMV quadratic estimators (or iterative reconstruction in brackets). CMB-S4 deep uses polarization-only reconstruction, and assumes a fractional sky coverage $f_{\textrm{sky}}=0.05$ with no foregrounds.}}
\label{tab:fish_opt}
\end{table}

When only considering a set of two unique tracers for template construction, the trends seen in the bispectra forecasts (and correlation coefficients) are also seen in these cross-correlation forecasts. Optimally constructed templates $\hat{\omega}^{\{g,I\}}$ and $\hat{\omega}^{\{\kappa,g\}}$ produce the most detectable cross-spectra, with $>4.5\sigma$ achievable for SO at goal sensitivity. When looking ahead to CMB-S4, $\hat{\omega}^{\{\kappa,g\}}$ gets the greatest boost due to the lower reconstruction levels in both $\hat{\kappa}$ and $\hat{\omega}$, and consequently is detectable at $12.1\sigma$-$13.3\sigma$. However, $\hat{\omega}^{\{g,I\}}$ only benefits from the noise reduction in $\hat{\omega}$, and so is slightly less enhanced at $9.3\sigma$-$9.8\sigma$. This is also seen for CMB-S4 deep patch with iterative reconstruction, in which $\hat{\omega}^{\{\kappa,g\}}$ and $\hat{\omega}^{\{g,I\}}$ are forecast at $20.4\sigma$ and $11.6$ respectively. The template $\hat{\omega}^{\{\kappa,I\}}$ is only detectable at $3.7\sigma$ in the best case for CMB-S4 deep patch with iterative reconstruction.

\begin{figure}[t]
 	\includegraphics[width=\linewidth]{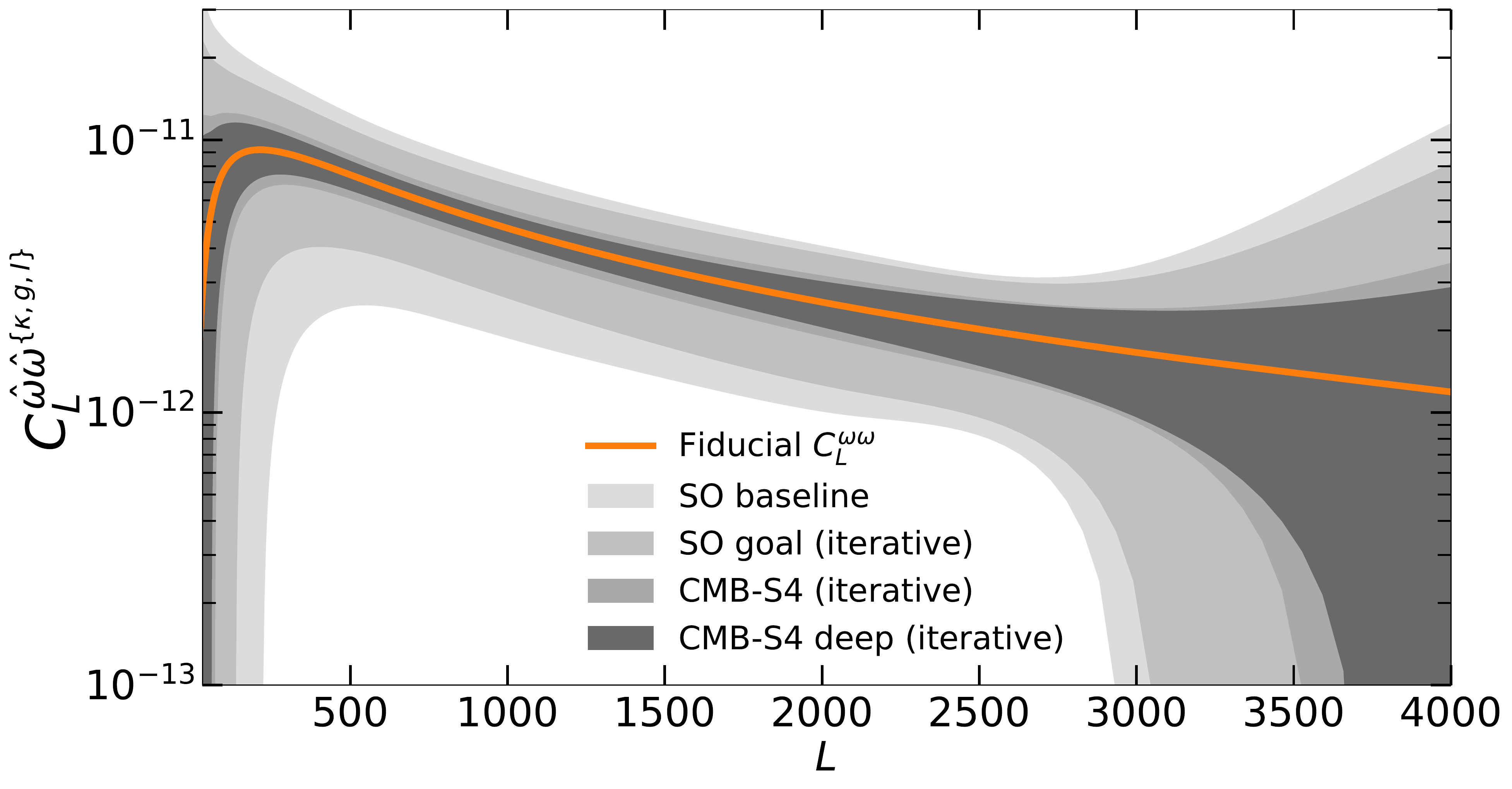}
	\caption{\small{The reconstruction error forecast on the template-rotation cross-spectrum using Eq.~\eqref{eq:cross_var}. Error bands are shown for 17 logarithmically separated bins in $L$, and interpolated to produce smooth contours. Different shades represent different experimental configurations, and the fiducial post-Born rotation is displayed in solid orange. The CMB-S4 deep setup use polarization-only reconstruction, and has a sky coverage of just $5\%$; the increase in cosmic variance with smaller sky area is compensated on all scales by reduced rotation reconstruction noise due the lower instrumental noise, with additional gains from lower convergence tracer reconstruction noise and the limiting assumption of no foregrounds on the deep patch. }}
	\label{fig:omega_cross_binned}
\end{figure}

The variance of the cross-spectrum for a bin centred at $L$ with width $\Delta L$ is given by
\begin{equation}\label{eq:cross_var}
\sigma^2_L=\frac{C^{\omega\omega}_LN^{\omega}_LF_L^{-1}}{\Delta L(2L+1)f_{\textrm{sky}}}.
\end{equation}
The $1\sigma$ error forecasts for the various experimental configurations are shown in Fig.~\ref{fig:omega_cross_binned}, assuming the optimal template estimator, $\hat{\omega}^{\{\kappa,g,I\}}$. The spectrum is constrained away from zero for all noise levels in the multipole range $200<L<3000$. At large scales the lack of modes (cosmic variance) dominates as is shown by the relatively larger error bars. Separately, on the small scales the CMB reconstruction noise starts to significantly dominate over the signal at $L\sim3000$, also seen in Fig. \ref{fig:omega_lss_ps}. While the amplitude of the template spectrum depends on the galaxy bias, the detectability forecasts are largely independent of the bias due to the high $S/N$ of $\hat{g}$. As long as the bias can be measured empirically (e.g. using cross-correlation with lensing convergence), the rotation template amplitude can be normalized empirically. Uncertainties in modelling of the galaxy bias (and potentially redshift distribution) only become relevant if the template amplitude is used for cosmological inference; however, the parameters are likely to be well constrained by convergence-density spectra compared to the uncertainty levels that would be relevant for the rotation.

There are a few avenues for improving these rotation forecasts further. For example, other density tracers could be used as input to the rotation template. Cosmic shear is the obvious candidate; although there would be significant overlap of redshift sensitivity with $\hat{g}$, cosmic shear is a more direct probe of the lensing effect. Another approach could be to use the full tomography of galaxy surveys (using galaxy counts and/or cosmic shear), giving a closer match to the ideal template. However, we only considered one large effective bin for $\hat{g}$ here, as we found improvements with binning to be modest. Finally, using external tracers to improve the delensing in the iterative reconstruction can reduce the reconstruction noise further. For CMB-S4 deep, including $\hat{g}$ and $\hat{I}$ in the iterative reconstruction potentially gives an additional $\sim2\sigma$ detection significance. However, the improvement is less apparent for the wide experimental setups. Using external tracers in the delensing may also complicate the cross-spectrum bias, since $\hat\omega^{\textrm{tem}}$ is constructed from the same (or correlated) tracers.

\section{Cross-spectrum Bias}\label{sec:bias}
The rotation-template cross spectrum is in general a biased estimator of the post-Born rotation power spectrum

\begin{equation}\label{eq:cross_expectation}
\left\langle C^{\hat{\omega}\hat{\omega}^{\textrm{tem}}}\right \rangle = C^{\omega\omega} + \textrm{bias}.
\end{equation}
The bias is non-zero even when there is no deflection, so for accurate measurement of lensing rotation the bias must be accounted for. Large biases could potentially degrade the Fisher forecasts of the previous section. Here we explore the leading Gaussian contributions to this cross signal bias, and show that even configurations with bias of similar amplitude to $C^{\omega\omega}$ do not affect the $S/N$.

\subsection{Template-lensing Bias}
CMB field rotation is reconstructed from pairs of lensed CMB fields, $\tilde{X},\tilde{Y}\in \{\tilde{T},\tilde{E},\tilde{B}\}$, as shown in  Eq.~\eqref{eq:omega_qe}. Likewise, the rotation template Eq.~\eqref{eq:flat_omega_QE_explicit} involves pairs of LSS tracers, e.g. $\hat{a},\hat{b}\in \{\hat{\kappa},\hat{g},\hat{I}\}$. Therefore, the full expectation of the cross-spectrum, i.e. the LHS of Eq.~\eqref{eq:cross_expectation}, is proportional to the trispectrum $\langle \tilde{X}\tilde{Y}\hat{a} \hat{b}\rangle$.

As described in Section~\ref{sec:background}, $\tilde{X}$ is related to the unlensed field, $X$, by the deflection field given in Eq.~\eqref{eq:alpha}. Consider the series expansion of the lensed field at some point in sky in terms of the deflection field (ignoring curl terms here as they are subdominant to the lensing potential)
\begin{align}\label{eq:lensed_field_expansion}
\begin{split}
\tilde{X} &= X + \delta_{\phi} X + \delta_{\phi}^2X+...\\
&= X +  \nabla^i\phi  \nabla_iX +  \frac{1}{2}\nabla^i\phi\nabla^j\phi  \nabla_i\nabla_jX +...
\end{split}
\end{align}
Therefore, from perturbative expansion of the trispectrum, the leading\footnote{We postpone discussion till Section \ref{sec:bias_add} of additional contributions that arise when one or both of $\hat a$, $\hat b$ are also derived from a quadratic estimator.} terms present in the absence of rotation are
\begin{align}\label{eq:bias_expand}
\begin{split}
\langle \tilde{X}\tilde{Y}\hat{a} \hat{b}\rangle_{\textrm{bias}} =  \langle \delta_{\phi}^2 XY\hat{a}\hat{b}\rangle + \langle X\delta_{\phi}^2Y\hat{a}\hat{b}\rangle \\
+ \langle \delta_{\phi} X\delta_{\phi} Y\hat{a}\hat{b}\rangle + \mathcal{O}(\Psi^6).
\end{split}
\end{align}
These three terms contribute to the bias at the $N^{(2)}$ level\footnote{Following the literature convention to label lensing bias terms by their order in $C^{\phi\phi}$ about the Gaussian result, which is equivalent to labelling by $\mathcal{O}(\Psi^2)$.} and are similar to the $N^{(3/2)}$ bias terms in Ref.~\cite[]{Fabbian:2019tik} which arise from a non-negligible convergence bispectra in the CMB lensing--LSS cross correlation. However, here we are still simply assuming Gaussianity of all density tracers (including $\kappa$), so this bias is not caused by a LSS/lensing bispectrum. We can see that each term is actually proportional to a density trispectrum $\langle\phi\phi \hat{a}\hat{b}\rangle$, which we assume is dominated by the disconnected part. Alternatively, as we are using pairs of tracers to estimate the lensing rotation, so the rotation template is non-Gaussian, the bias could be also be considered as coming from template-lensing bispectra $\langle\phi\phi \hat{\omega}^{\textrm{tem}}\rangle$. We choose to follow the notational conventions of Ref.~\cite[]{Bohm:2016gzt} and label the first two terms of Eq.~\eqref{eq:bias_expand} as $N^{(2)}_{\textrm{C1}}$, and the last as $N^{(2)}_{\textrm{A1}}$.

Assuming the unlensed CMB and late-time density tracers are independent (and Gaussianity of all the fields), using the perturbative expansion of the lensed CMB fields gives the analytic results
\begin{align}\label{eq:NA1}
\begin{split}
    N^{(2)}_{\mathrm{A_1}}(L)=-\frac{L^2}{2}A^{\Omega}_L\int \frac{d^2\boldsymbol{l}d^2\boldsymbol{L}_1}{(2\pi)^4}b^{\hat{\omega}^{\textrm{tem}}\phi\phi}_{(-\boldsymbol{L})\boldsymbol{L}_1\boldsymbol{L}_2}(\boldsymbol{l}'' \cdot\boldsymbol{L}_1)(\boldsymbol{l}'' \cdot\boldsymbol{L}_2)\\
\sum_{X,Y}g^{\Omega}_{XY}(\boldsymbol{l},\boldsymbol{L})C^{\bar{X}\bar{Y}}_{l''}h_X(\boldsymbol{l}'',\boldsymbol{l})h_Y(\boldsymbol{l}'',\boldsymbol{l}'),
\end{split}
\end{align}
and
\begin{align}\label{eq:NC1}
\begin{split}
    N^{(2)}_{\mathrm{C_1}}(L)=\frac{L^2}{4}A^{\Omega}_L\int \frac{d^2\boldsymbol{l}d^2\boldsymbol{L}_1}{(2\pi)^4}b^{\hat{\omega}^{\textrm{tem}}\phi\phi}_{(-\boldsymbol{L})\boldsymbol{L}_1\boldsymbol{L}_2}(\boldsymbol{l} \cdot\boldsymbol{L}_1)(\boldsymbol{l} \cdot\boldsymbol{L}_2)\\
    \sum_{X,Y}\bigg[g^{\Omega}_{XY}(\boldsymbol{l},\boldsymbol{L})C^{X\bar{Y}}_{l}h_{Y}(\boldsymbol{l}, \boldsymbol{l}')\\
 + g^{\Omega}_{YX}(\boldsymbol{l},\boldsymbol{L})C^{\bar{X}Y}_{l}h_{X}(\boldsymbol{l}, \boldsymbol{l}')\bigg],
\end{split}
\end{align}
for $\boldsymbol{L}_2=\boldsymbol{L}-\boldsymbol{L}_1$, $\boldsymbol{l}'=\boldsymbol{L}-\boldsymbol{l}$, and
$\boldsymbol{l}''\equiv\boldsymbol{L}_1-\boldsymbol{l}$. Here, $\bar{T}=T$, $\bar{E}=E$ and $\bar{B}=E$, and the geometric functions, $h$, come from the response function definitions in Table \ref{tab:resps}. In Eqs.~\eqref{eq:NA1} and \eqref{eq:NC1} we follow the approximate non-perturbative result of Ref.~\cite[]{Fabbian:2019tik} by replacing unlensed CMB spectra with the corresponding lensed gradient spectra, $C^{XY}\rightarrow \tilde{C}^{X\nabla Y}$. We obtain the template rotation bispectra by contraction of the lensing fields with Eq.~\eqref{eq:flat_omega_QE_explicit}, giving
\begin{equation}\label{eq:mixed_bi}
    \begin{split}
        b^{\hat{\omega}^{\textrm{tem}}\phi\phi}_{(-\boldsymbol{L})\boldsymbol{L}_1\boldsymbol{L}_2} &= \frac{4}{L_1^2L_2^2}b^{\hat{\omega}^{\textrm{tem}}\kappa\kappa}_{(-\boldsymbol{L})\boldsymbol{L}_1\boldsymbol{L}_2}\\
        &\equiv\frac{4F_L^{-1}}{L_1^2L_2^2}b^{\omega ij}_{(-\boldsymbol{L})\boldsymbol{L}_1\boldsymbol{L}_2}(\boldsymbol{C}^{-1}_{\textrm{LSS}})^{ip}_{L_1}(\boldsymbol{C}^{-1}_{\textrm{LSS}})^{jq}_{L_2}C^{p\kappa}_{L_1}C^{q\kappa}_{L_2}.
    \end{split}
\end{equation}
With a non-zero fiducial $b^{\omega ij}$ in the weights, the template-convergence bispectra are non-zero even if the data rotation is zero and all the fields are Gaussian.

\begin{figure}[t]
 	\includegraphics[width=\linewidth]{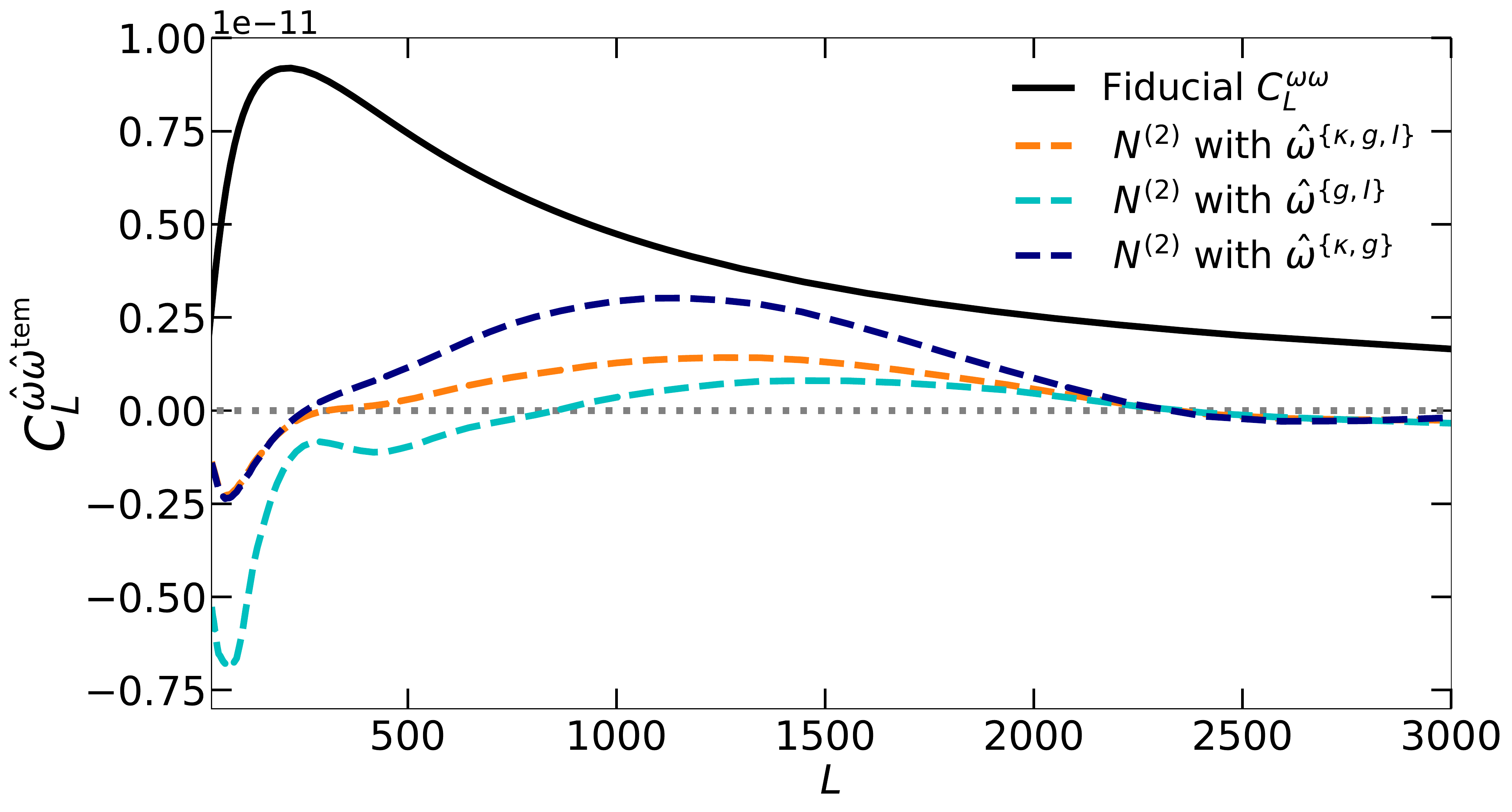}
	\caption{\small{Analytic predictions for the $N^{(2)}_{\mathrm{A1}}+N^{(2)}_{\mathrm{C1}}$ bias (dashed lines) on the rotation-template cross-spectrum $C^{\hat{\omega}\hat{\omega}^{\textrm{tem}}}$ for different templates at SO goal sensitivity, assuming Gaussian fields. The lensing observables, $\hat{\omega}$ and $\hat{\kappa}$, are reconstructed with the GMV estimator. The curves can be compared to the fiducial post-Born rotation power spectrum in solid black, and the null dotted line.
Contributions from the contractions entering the $\hat\kappa$ quadratic estimator are neglected here, apart from the linear $\kappa$ field itself.
}}\label{fig:bias_results}
\end{figure}

Fig.~\ref{fig:bias_results} shows the total bias from all the $N^{(2)}$ terms for different rotation templates at SO goal sensitivity. Almost all of the analytic predictions for the bias are smaller than the expected fiducial rotation signal at all multipoles. The smallest biases across most scales are for the optimal template, $\hat{\omega}^{\{\kappa,g,I\}}$, and the convergence-free template, $\hat{\omega}^{\{g,I\}}$. 
Here we have not included the bias from $\hat{\omega}^{\{\kappa,I\}}$, as the result is around an order of magnitude larger than the other configurations. As shown in Section~\ref{sec:fisher}, $\hat{\omega}^{\{\kappa,I\}}$ is the noisiest rotation template, and hence the least interesting to consider.

\subsection{Additional biases with $\hat{\kappa}$}\label{sec:bias_add}
The $N^{(2)}$ bias terms Eq.~\eqref{eq:NA1} and Eq.~\eqref{eq:NC1}, represent the leading biases only under the assumption that $\hat{X}$ and $\hat{a}$ are independent. This is most untrue for $\hat{a}=\hat{\kappa}$, which is reconstructed from pairs of lensed CMB maps, $\hat{\kappa}\propto \tilde{X}\tilde{Y}$. Therefore, the CMB fields used to reconstruct $\hat{\omega}$ will form contractions with the fields used to reconstruct $\hat{\kappa}$, and induce extra biases if $\hat{\kappa}$ is included as a density tracer. There are several configurations in which these additional biases are generated below $N^{(2)}$ level.

Consider the different tracer combinations that go into the construction of $\hat{\omega}^{\textrm{tem}}$. If $\hat{\kappa}$ is used as one of the input tracers, there will be a term in Eq.~\eqref{eq:flat_omega_QE_explicit} such that $\hat{a}^i\hat{a}^j=\hat{\kappa}\hat{\kappa}$. For this term, the cross-spectrum is related to a 6-point function of CMB maps $\langle\hat{\omega}\hat{\omega}^{\textrm{tem}}\rangle\propto\langle\tilde{X}\tilde{Y}\tilde{X}\tilde{Y}\tilde{X}\tilde{Y}\rangle$. 
This has a fully disconnected contribution at $N^{(0)}$ level, leaving 8 terms after Wick expansion,
the combination of which produce the total $\hat{\kappa}\hat{\kappa}$ leading order bias
\begin{multline}\label{eq:N0_add}
   N^{(0)}_{\hat{\kappa}\hat{\kappa}}(L)=2L^2A^{\Omega}_L\int \frac{d^2\vl d^2\vL_1}{(2\pi)^4} \beta(\boldsymbol{L},\boldsymbol{L}_1)g^{\Omega}_{ij}(\boldsymbol{l},\boldsymbol{L})\\
g^{\phi}_{pq}(\boldsymbol{l},\boldsymbol{L}_1)g^{\phi}_{rs}(\boldsymbol{l'},\boldsymbol{L}_2)
(\tilde{\boldsymbol{C}}_{\textrm{CMB}})^{ip}_{l}(\tilde{\boldsymbol{C}}_{\textrm{CMB}})^{jr}_{l'}(\tilde{\boldsymbol{C}}_{\textrm{CMB}})^{qs}_{l''},
\end{multline}
where the LSS mode coupling function for $N^{(0)}_{\hat{\kappa}\hat{\kappa}}$ is defined by
\begin{equation}
\beta(\boldsymbol{L},\boldsymbol{L}_1) \equiv \frac{L_1^2L_2^2}{4}A^{\phi}_{L_1}A^{\phi}_{L_2}
F_L^{-1}b^{\omega ij}_{(-\boldsymbol{L})\boldsymbol{L}_1\boldsymbol{L}_2}(\boldsymbol{C}^{-1}_{\textrm{LSS}})^{i\kappa}_{L_1}(\boldsymbol{C}^{-1}_{\textrm{LSS}})^{j\kappa}_{L_2}.
\end{equation}

Now consider all the different tracer pairs in which $\hat{\kappa}$ accounts for only one leg of the template estimator, $\hat{a}^i\hat{a}^j\in\{\hat{\kappa}g,\hat{\kappa}I,g\hat{\kappa},I\hat{\kappa}\}$. For each of these cases, the cross-spectrum is now a CMB-tracer 5-point function $\langle\hat{\omega}\hat{\omega}^{\textrm{tem}}\rangle\propto\langle\tilde{X}\tilde{Y}\tilde{X}\tilde{Y}\hat{a}^j\rangle$.
The leading bias in this case is at $N^{(1)}$ level, and is best captured non-perturbatively using definition of the lensing potential response functions (equivalent to Eq.~\eqref{eq:resp} for the $\Omega$ responses; explicit $f^{\phi}$ functions given in Table \ref{tab:resps_kappa}). There are then 4 distinct terms that contribute to the one-$\hat{\kappa}$ bias at leading order. Using the symmetry property of the GMV weights, $g_{ij}(\boldsymbol{l},\boldsymbol{L})=g_{ji}(\boldsymbol{L}-\boldsymbol{l},\boldsymbol{L})$, this bias reduces to
\begin{multline}\label{eq:N1_add}
   N^{(1)}_{\hat{\kappa}}(L)=2L^2A^{\Omega}_L\int \frac{d^2\vl d^2\vL_1}{(2\pi)^4} \alpha(\boldsymbol{L},\boldsymbol{L}_1)\\
g^{\Omega}_{ij}(\boldsymbol{l},\boldsymbol{L})g^{\phi}_{pq}(\boldsymbol{l},\boldsymbol{L}_1)(\tilde{\boldsymbol{C}}_{\textrm{CMB}})^{ip}_{l}f^{\phi}_{jq}(\boldsymbol{l}',\boldsymbol{L}_2),
\end{multline}
complete with definition of the corresponding LSS function
\begin{equation}\label{eq:N1_alpha}
\alpha(\boldsymbol{L},\boldsymbol{L}_1) \equiv \frac{L_1^2}{{L_2^2}}A^{\phi}_{L_1}F_L^{-1}b^{\omega ij}_{(-\boldsymbol{L})\boldsymbol{L}_1\boldsymbol{L}_2}(\boldsymbol{C}^{-1}_{\textrm{LSS}})^{i\kappa}_{L_1}(\boldsymbol{C}^{-1}_{\textrm{LSS}})^{jq}_{L_2}C^{\kappa q}_{L_2}.
\end{equation}

\begin{figure}[t]
 	\includegraphics[width=\linewidth]{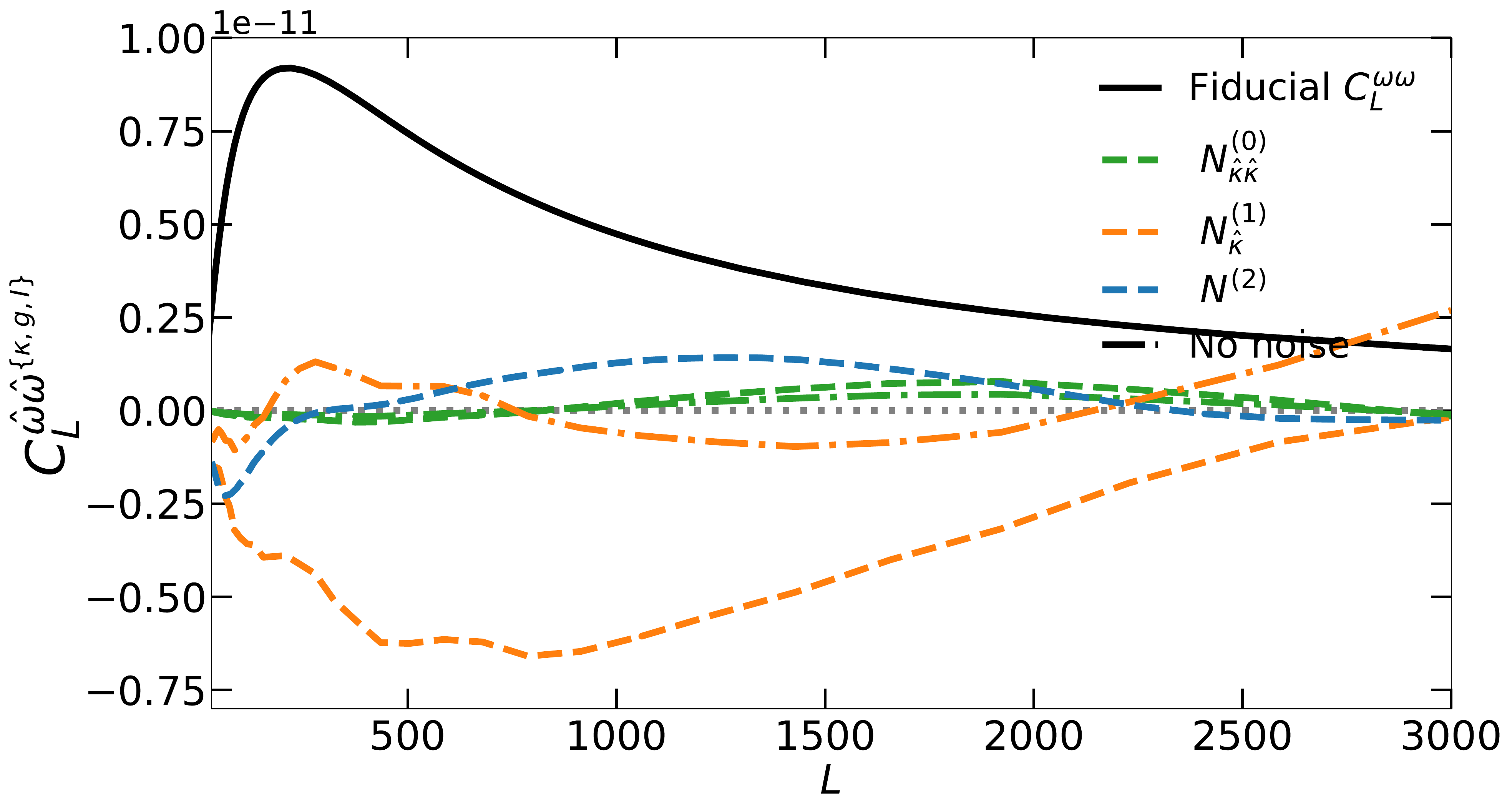}
	\caption{\small{Analytic predictions for all biases considered (dashed lines) on the rotation-template cross-spectrum $C^{\hat{\omega}\hat{\omega}^{\textrm{tem}}}$ at SO goal sensitivity, assuming Gaussian tracer and CMB fields. The lensing observables, $\hat{\omega}$ and $\hat{\kappa}$, are reconstructed with the GMV estimator. The curves can be compared to the fiducial post-Born rotation power spectrum in solid black, and the null dotted line. Contributions from contractions involving the $\hat\kappa$ quadratic estimator are shown by the green and orange lines for $N^{(0)}_{\hat{\kappa}\hat{\kappa}}$ and $N^{(1)}_{\hat{\kappa}}$ respectively. If $\hat{\omega}$ and $\hat{\kappa}$ are measured from independent CMB maps then $N^{(0)}_{\hat{\kappa}\hat{\kappa}}$ and $N^{(1)}_{\hat{\kappa}}$ lose the contribution from CMB map level noise and reduce to the corresponding dash-dotted lines.}}\label{fig:bias_all_levels}
\end{figure}

Fig.~\ref{fig:bias_all_levels} shows the biases on the rotation cross-spectrum with the optimal template $\hat{\omega}^{\{\kappa,g,I\}}$ at SO goal sensitivity. The $N^{(1)}_{\hat{\kappa}}$ term dominates and is of comparable magnitude to the signal over the multipoles of interest, $500\lesssim L\lesssim2000$. The observed CMB power spectra $\tilde{\boldsymbol{C}}_{\textrm{CMB}}$ in Eq.~\eqref{eq:N1_add} determine the size of this bias. The bias can therefore be reduced by using CMB measurements with no noise correlations, for example if $\hat{\omega}$ and $\hat{\kappa}$ are reconstructed with different sets of CMB maps so that there are no noise correlations between maps used for $\hat{\omega}$ and maps used for $\hat{\kappa}$.
The dash-dot lines in Fig.~\ref{fig:bias_all_levels} show the size and shape of $N^{(1)}_{\hat{\kappa}}$ is significantly reduced if the maps have no noise; the remaining $N^{(1)}_{\hat{\kappa}}$ is then comparable in magnitude to $N^{(2)}$. Similarly, $N^{(0)}_{\hat{\kappa}\hat{\kappa}}$ moves closer to the null line on all scales for this case.

There are several possible ways to mitigate the contribution of these additional $\hat{\kappa}$ bias signals. One could use $\hat{\kappa}$ and $\hat{\omega}$ reconstructed from maps split up over different time periods \cite{Madhavacheril:2020ido,ACT:2023dou} to get rid of the noise contributions and reduce the biases back to dash-dot lines in Fig. \ref{fig:bias_all_levels} (in the approximation that any splits do not affect map level noise). Equivalently, they could be reconstructed from different surveys or detectors. Alternatively, $\hat{\kappa}$ could be reconstructed from T only, while $\hat{\omega}$ is reconstructed with a polarization-only estimator, then only the $\tilde{C}^{TE}_l$ configuration in $(\tilde{\boldsymbol{C}}_{\textrm{CMB}})^{XY}_l$ survives in Eq.~\eqref{eq:N1_alpha}, potentially resulting in another reduction to the bias magnitude. To avoid the $N^{(0)}$ or $N^{(1)}$ entirely, one could simply construct a convergence-free $\hat{\omega}^{\textrm{tem}}$, for example by only using $\hat{a}\in \{\hat{g},\hat{I}\}$. We have already shown that such a template is detectable with high significance in the near future, and its bias is small compared to the fiducial rotation signal (Fig.~\ref{fig:bias_results}).

We have not included biases for lensing reconstruction with iterative estimators in our investigation, as the analytical form of iterative bias terms requires careful consideration. Additionally, the biases considered here are for full GMV reconstruction only, and do not directly apply to the CMB-S4 deep experimental setup which uses polarization-only reconstruction. Hence, analysis of biases appropriate to CMB-S4 deep is left to future work. Possible bias contributions from correlations between CMB foreground residuals and the LSS tracers have also been neglected. Finally, it is worth re-emphasising that only Gaussian biases are considered here, the investigation of contamination from higher order statistics is left to future work.

\begin{figure*}[t]
 	\includegraphics[width=\linewidth]{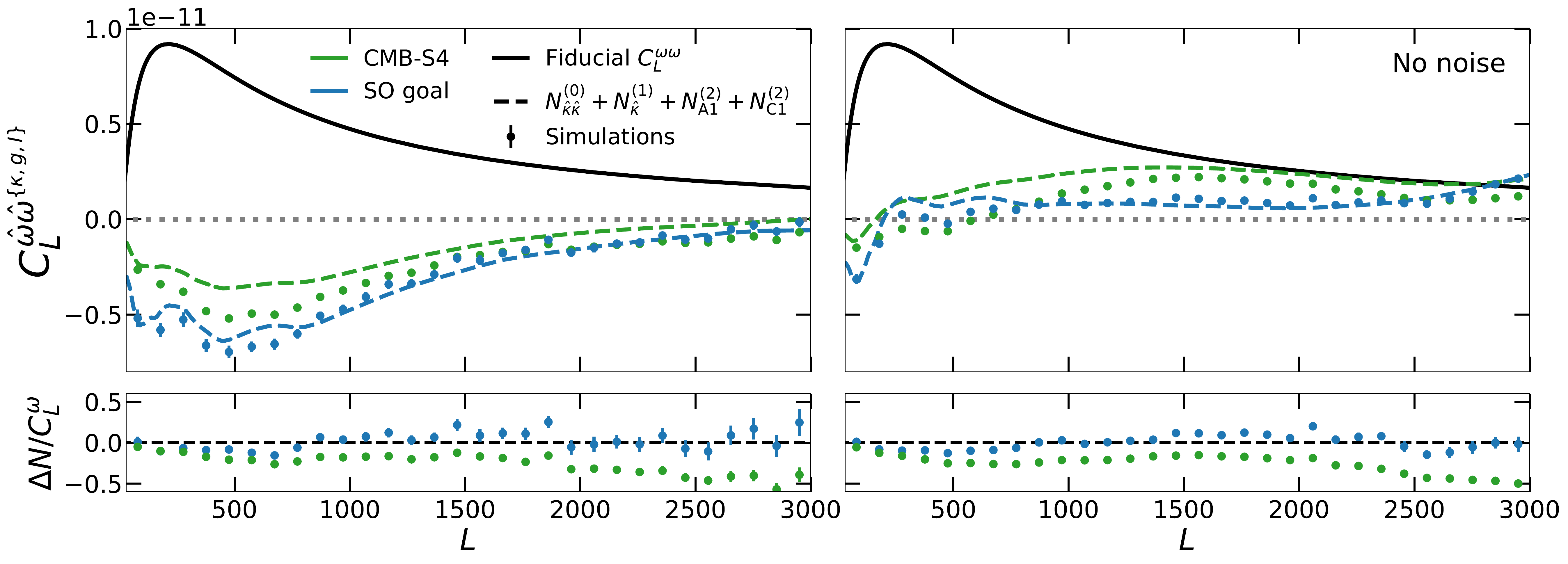}
	\caption{\small{The total Gaussian bias on the optimal rotation-template cross-spectrum $C^{\hat{\omega}\hat{\omega}^{\{\kappa,g,I\}}}$ for SO at goal sensitivity (blue) and CMB-S4 (green), assuming $\hat{\omega}$ and $\hat{\kappa}$ are reconstructed with the GMV estimator. Dashed lines show the combined analytic predictions for $N^{(0)}_{\hat{\kappa}\hat{\kappa}}$, $N^{(1)}_{\hat{\kappa}}$, $N^{(2)}_{\textrm{A1}}$ and $N^{(2)}_{\textrm{C1}}$. Points show the mean bias from 100 Monte Carlo simulations for 30 linearly-spaced bins in $L$, with error bars representing standard error in the mean. The results can be compared to the fiducial post-Born rotation power spectrum in solid black, and the null dotted line. {\it Left panel:} Contributions to the bias from CMB map level noise are included in the theoretical curves and simulated results. This represents the case in which $\hat{\omega}$ and $\hat{\kappa}$ are reconstructed from the same CMB maps. {\it Right panel:} Same as left panel with exception that  $\hat{\omega}$ and $\hat{\kappa}$ are now measured using CMB maps with independent noise. Therefore noise is not included in the simulations at map level, and similarly noise contributions are not included in the theory lines. {\it Bottom panels:} Residuals of the total simulated Gaussian bias normalized by fiducial post-Born $C^{\omega\omega}_L$ show qualitative agreement between theory and simulations for SO goal. Unaccounted contributions from $N^{(2)}_{\hat{\kappa}}$ explain the CMB-S4 discrepancy.}}\label{fig:bias_final_big}
\end{figure*}

\subsection{Testing using Simulations}\label{sec:sims}

We use Gaussian Monte Carlo simulations to test the analytically-derived bias terms from the previous two sections. In this section, all lensing reconstruction is achieved via the full GMV estimator for SO goal and CMB-S4 like experiments, and only the optimal rotation template $\hat{\omega}^{\{\kappa,g,I\}}$ is considered.

For no fiducial rotation in the simulations, the cross spectrum is purely bias:
\begin{equation}
\langle C^{\hat{\omega}\hat{\omega}^{\textrm{tem}}}\rangle_{\omega=0}=\textrm{bias} = N^{(0)}+N^{(1)}+N^{(2)}+\mathcal{O}(\Psi^{6}).
\end{equation}
We therefore reconstruct $\hat{\omega}$ and $\hat{\kappa}$ from flat-sky lensed CMB simulations in which $\omega=0$ in the maps. The code {\sc LensIt} \cite[]{Carron:2017mqf} performs the lensing of the CMB maps and the reconstruction. We use the same experimental configurations as the Fisher forecasts, and the maps are again cut at $30\leq L\leq3000$ for $T$, and $30\leq L\leq5000$ for $E$ and $B$ to maintain consistency.

In preparation for the template estimator, a $\kappa$ realization is used to generate Gaussian maps of $\hat{g}$ and $\hat{I}$ via the Cholesky decomposition of their covariances, defined by Eq.~\eqref{eq:kappa_ps}, Eq.~\eqref{eq:density_tracer_ps}, and Eq.~\eqref{eq:kappa_tracer_ps}.
This provides the necessary cross-correlations between tracers, and dependence between $\hat{\omega}$ and the tracer maps. The optimal rotation template map, $\hat{\omega}^{\{\kappa,g,I\}}$, is then constructed using the fast configuration-space method of Eq.~\eqref{eq:omega_qe_split_ft}, with the generated tracer maps plus the lensing reconstructed $\hat{\kappa}$ as inputs.

All maps are generated to have sky coverage of $\sim41,000\, {\rm deg}^2$ corresponding to an $f_{\rm sky}=1$. There are $(4096)^2$ pixels in each sky patch, so there are modes resolved up to $L\sim5100$. We apply cuts on the tracer maps at $30\leq L\leq3000$ (with the exception of the CIB which has the same stringent cuts of $110\leq L\leq2000$) and all modes within this range are fully resolved. We generate 100 sets of simulations.

We first consider the case of independent CMB noise realizations in the $\hat{\omega}$ and $\hat{\kappa}$ reconstructions, so that no contributions to $N^{(0)}_{\hat{\kappa}\hat{\kappa}}$ or $N^{(1)}_{\hat{\kappa}}$ come from map-level noise
($N^{(2)}_{\textrm{A1}}$ and $N^{(2)}_{\textrm{C1}}$ remain unaffected by the map level noise). Therefore, we do not add noise to the CMB or LSS maps as it would only increase the scatter of the results\footnote{Note the $\hat\omega$ and $\hat\kappa$ maps still have reconstruction noise from CMB fluctuations.}.
However, noise power is included in the filters used for the GMV reconstruction, i.e. in the inverse covariance weights of Eq.~\eqref{eq:gmv_weight}, and the inverse filtered LSS maps in Eq.~\eqref{eq:flat_omega_QE_explicit}. The total biases present in these noiseless simulations are represented by the points in right panel of Fig.~\ref{fig:bias_final_big}. The bottom panel shows the residuals between simulation and the analytic prediction, normalized by the fiducial post-Born rotation spectrum. For SO at goal sensitivity there is good qualitative agreement that the shape and size of the total bias is almost entirely described by the analytic predictions. This contrasts with CMB-S4; while there is agreement with the overall shape of the bias, there is an offset where the analytic prediction overestimates the size of the bias.

We also test the total bias with map level noise contributions included. In this case, noise is added to the CMB maps used in the reconstruction of $\hat{\omega}$ and $\hat{\kappa}$. Here we also add noise to the tracer maps. The inclusion of tracer noise does not generate new biases; we have consistently assumed noise is independent between LSS fields throughout. However, it does increase the error bars on the simulated bias signal. This enables comparison between the scatter on the simulated bias with our forecast 1-$\sigma$ error on the rotation cross-spectrum in Fig.~\ref{fig:omega_cross_binned} to check whether the variance on the bias affects the Fisher forecasts.

The left panel of Fig.~\ref{fig:bias_final_big} shows the total simulated bias on the rotation cross-spectrum. Again, the the shape and size of the total bias in the SO goal setup is well described by the analytic predictions. However, the CMB-S4 analytic biases under-predict the total bias from the simulations, though the shapes are qualitatively consistent. The mischaracterization of CMB-S4 biases in both cases considered would constitute a $\sim20$--$50\%$ residual offset on measurement of the fiducial spectrum if reliant purely on the analytic predictions of $N^{(0)}_{\hat{\kappa}\hat{\kappa}}$, $N^{(1)}_{\hat{\kappa}}$, $N^{(2)}_{\textrm{A1}}$ and $N^{(2)}_{\textrm{C1}}$. We find the discrepancy to be consistent with a missing theoretical bias contribution from $N^{(2)}_{\hat{\kappa}}$ like terms which are perhaps more prevalent for CMB-S4 due to lower noise on the polarization. This is explored further in Appendix \ref{app:bias} along with further validation tests for individual bias terms. We leave explicit derivation of $N^{(2)}_{\hat{\kappa}}$ to future work.

Finally, we find the simulations to be in excellent agreement with the size of the forecast error-bars on the rotation cross-spectrum in Fig.~\ref{fig:omega_cross_binned} despite inclusion of the biases shown in Fig. \ref{fig:bias_final_big}. Hence, for near-future CMB experiments, provided any bias can be reasonably estimated from simulations, subtraction of these Gaussian biases from the measured cross-spectrum is sufficient to probe post-Born rotation power without compromising $S/N$.

\section{Conclusions}\label{sec:conclusion}

We have shown lensing rotation to be detectable at high significance by combining a lensing reconstruction measurement with a template constructed from the lensing convergence and/or large-scale structure tracers. Detection of this rotation signal will give a powerful internal consistency check of the cosmological model, data modelling and systematics.  Since the signal is detectable at high significance, the rotation power spectrum shape can also be recovered at some level.

The quadratic template for lensing rotation that we have constructed could also be used to `rotation-clean' the lensing reconstruction signal, allowing use of the cleaned signal as a more stringent null test or probe of non-standard cosmological signals. E-mode polarization observations combined with the rotation template could also be used to remove the rotation-induced B-mode polarization, in a similar way that E modes combined with large-scale structure tracers and lensing convergence can be used to delens the observed B-mode polarization produced by standard gradient lensing. This can potentially improve the sensitivity to B modes from primordial gravitational waves, though the rotation signal only becomes marginally significant for tensor-scalar ratios $r\alt 10^{-5}$~\cite{Lewis:2017ans}.

We have made a number of simplifying assumptions in this paper, that would have to be relaxed for application to more-realistic data. Although we have tested consistency with simple null-hypothesis simulations to check the size of the leading biases, future work should use more realistic non-Gaussian post-Born lensing simulations such as those in Ref.~\cite{Fabbian:2017wfp}.
This could test for correct recovery of a non-zero rotation signal, and quantify the importance of any additional biases. Future work could also quantify the biases that arise when using iterative lensing reconstruction.
If necessary, the biases can be reduced by using semi-independent subsets of the data, and by using different subsets it should also be possible to have multiple consistency checks between results from different rotation estimators.

The parity-odd nature of the rotation field makes it relatively insensitive to many systematics that plague other large-scale structure observations, however any systematics that can produce rotation-like effects would have to be assessed carefully, and additional real-life uncertainties could lead to larger variance. Handling uncertainties in the modelling of bias, and other effects relating observations to the underlying matter field, should also be investigated more thoroughly.

\begin{acknowledgements}
We thank Julien Carron for help with his \textsc{plancklens} and \textsc{LensIt} codes, and further thank Julien and Giulio Fabbian for insightful discussions. We also thank Toshiya Namikawa and Byeonghee Yu for providing details of their CIB modelling. MR is supported by a UK Science and Technology Facilities Council (STFC) studentship. AL was supported by the UK STFC grants ST/T000473/1 and ST/X001040/1. Parts of this research used resources of the National Energy Research Scientific Computing Center (NERSC), a U.S. Department of Energy Office of Science User Facility located at Lawrence Berkeley National Laboratory.
\end{acknowledgements}

\appendix
\section{Iterative estimators}\label{app:iters}
The lensing reconstruction noise from iterative lensing estimation can be estimated quite easily from a modification of the standard analytic result for $N_0$~\cite{Smith:2010gu,Carron:2017mqf,Legrand:2021qdu}. At each step of the iteration, the reconstruction is effectively performed on a sky that has been delensed using the current best estimator for the lensing field (which is always dominated by convergence, since the rotation is very small).
Instead of using the noise and lensed CMB power spectra to calculate the disconnected Gaussian $N_0$ variance, the iterative reconstruction noise can be evaluated by using partially delensed CMB power spectra instead. This reduces the reconstruction noise, primarily due to the reduced variance from the lensing-dominated B modes. The amount of delensing that is possible at each $L$ depends on the cross-correlation of the current lensing estimate $\hat\phi_i$ with the truth:
\begin{equation}
  \rho_i^2 \equiv \frac{(C^{\phi\hat{\phi}_i})^2}{C^{\hat\phi_i\hat\phi_i}C^{\phi\phi}} = \frac{C^{\phi\phi}}{C^{\hat\phi_i\hat\phi_i}} =
  \frac{C^{\phi\phi}}{C^{\phi\phi} + N_0^{(i)}},
\end{equation}
where we have assumed the estimator is unbiased in the sense that $C^{\hat\phi_i \phi} = C^{\phi\phi}$ and neglected $N^{(1)}$.
The lensing remaining after partial delensing has power spectrum $(1-\rho_i^2)C^{\phi\phi}$, which is what is used to calculate the partially delensed spectra. As iterations progress, the delensed power, and hence $N_0^{(i)}$, converge.
This noise estimation procedure differs slightly from the method usually used to estimate delensed B-mode noise using Wiener-filtered E modes~\cite{Smith:2010gu} (also different from Ref.~\cite{Hotinli:2021umk}), but in practice the forecast results are very similar.

In addition to internal delensing at each step, it is also possible to use one or more external tracers~\cite{Smith:2010gu,Sherwin:2015baa}. The optimal combination of the internal lensing estimator at each step, and an external estimate with cross-correlation, $\rho_{\rm ext}$, to the true lensing field, has a total cross-correlation given by~\cite{Sherwin:2015baa}
\begin{equation}
  \rho^2_{\rm tot} = \frac{ (1-\rho_{\rm ext}^2) C^{\phi\phi} + \rho_{\rm ext}^2 N_0}{(1-\rho_{\rm ext}^2)C^{\phi\phi} + N_0}.
\end{equation}
The joint iterative $N_0^{(i)}$ is then estimated simply by partially delensing using this combined correlation function.

\section{Density Tracers}\label{app:tracers}
\subsubsection{CMB Lensing Convergence}\label{sec:cmb_conv}

The lensing convergence was introduced in Section~\ref{sec:conv}, and described generally in the wider context of weak lensing. Here $\kappa$ refers to the CMB lensing convergence only.

\begin{figure}[t]
 	\includegraphics[width=\linewidth]{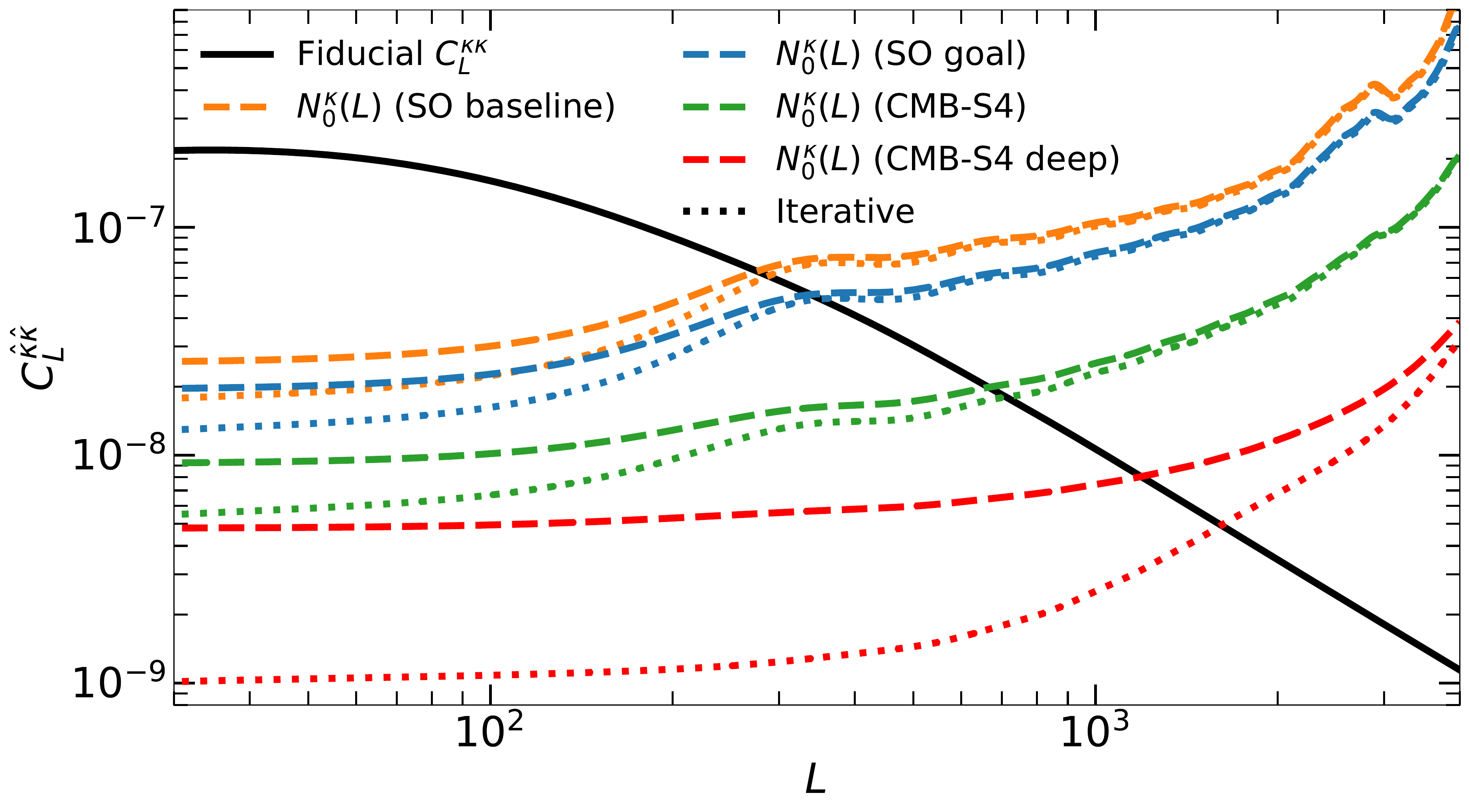}
	\caption{\small{The lowest-order CMB lensing reconstruction noise, $N^{\kappa}_0$, for GMV estimators are shown by the dashed lines for SO and CMB-S4 experimental configurations. Iterative reconstruction is shown in the dotted lines for the same experiments. CMB-S4 deep patch uses polarization-only reconstruction, and covers just $5\%$ of the sky and assumes no foregrounds. The fiducial lensing convergence power spectrum is shown in solid black.}}
	\label{fig:kappa_ps}
\end{figure}

The only noise term considered is the lowest order reconstruction bias $N^{\kappa}_L=N^{\kappa}_0(L)$ for either the GMV estimator (similar to the terms described in Section~\ref{sec:cmb_lensing} in the context of curl reconstruction), or iterative reconstruction (see Appendix \ref{app:iters}). These noise levels for $C^{\kappa\kappa}_L$ are illustrated in Fig. \ref{fig:kappa_ps} for the different experimental setups considered in this paper.

The window function for the convergence is written explicitly in Eq.~\eqref{eq:window_cmb}.

\subsubsection{Galaxy Clustering}
\begin{figure}[t]
 	\includegraphics[width=\linewidth]{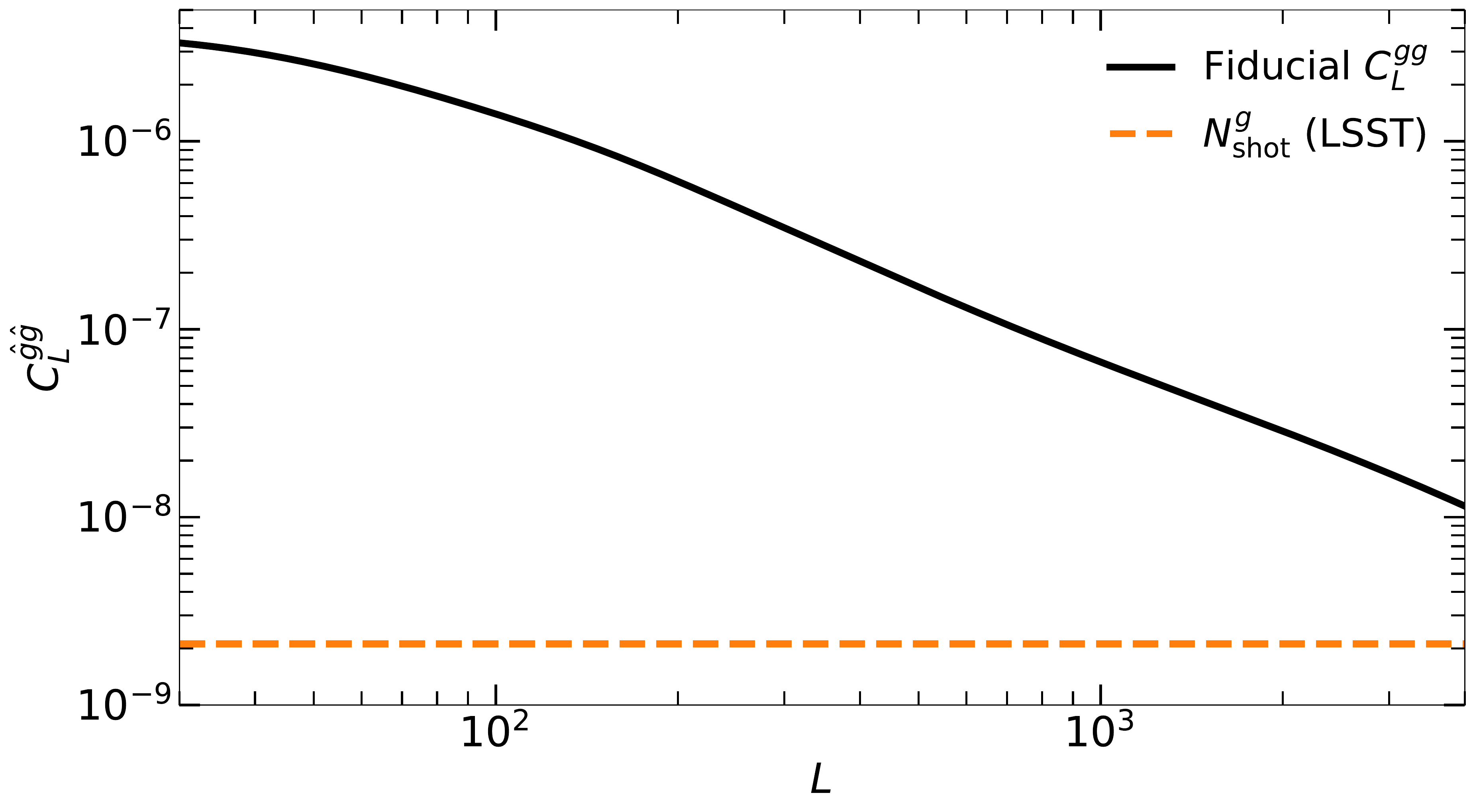}
	\caption{\small{Fiducial galaxy density power spectrum (solid line ) and the corresponding shot noise for an LSST-like survey (dashed line).}}
	\label{fig:gal_ps}
\end{figure}

Galaxies are a natural tracer of the matter distribution. They form within over-densities and are therefore a biased tracer of the LSS. Their window function is
\begin{equation}
   W_{g}(\chi)=b(z)n(z)H(z),
\end{equation}
for bias, $b$, and Hubble parameter, $H$, evaluated at redshift $z(\chi)$. The Hubble parameter is required to account for the change in integration variable from $z$ to $\chi$ in Eq.~\eqref{eq:density_tracer}. The redshift distribution is normalized as follows
\begin{equation}\label{eq:gal_distro1}
   n(z)=\frac{dN/dz}{\int dz'(dN/dz')}.
\end{equation}
We do not split the distribution into redshift bins, instead $g$ is considered for one effective redshift bin described by the LSST gold model \cite[]{LSSTScience:2009jmu}, which for a survey of limiting magnitude at 25.3 corresponding to $z_0=0.311$, is described by
\begin{equation}\label{eq:gal_distro2}
   \frac{dN}{dz}=\frac{1}{2z_0}\left(\frac{z}{z_0}\right)^{2}\exp\left(\frac{-z}{z_0}\right).
\end{equation}
The bias is modelled to linear order, $b(z)=1+0.84z$. We take a conservative forecast for the LSST gold sample as having a mean galaxy count of $\bar{n}=$ 40 galaxies/arcmin$^2$, and correspondingly the flat shot noise for galaxy counts is
\begin{equation}
    N^{g}_L = N^{g}_{\mathrm{shot}}=1/\bar{n},
\end{equation}
where $\bar{n}$ is converted to units of galaxies per steradian. The total galaxy power spectrum for the LSST gold sample is shown in Fig. \ref{fig:gal_ps}.

For simplicity, we do not consider redshift space distortions, magnification bias, or any other subdominant contributions to the galaxy clustering model.

\subsubsection{Cosmic Infrared Background}

\begin{figure}[t]
 	\includegraphics[width=\linewidth]{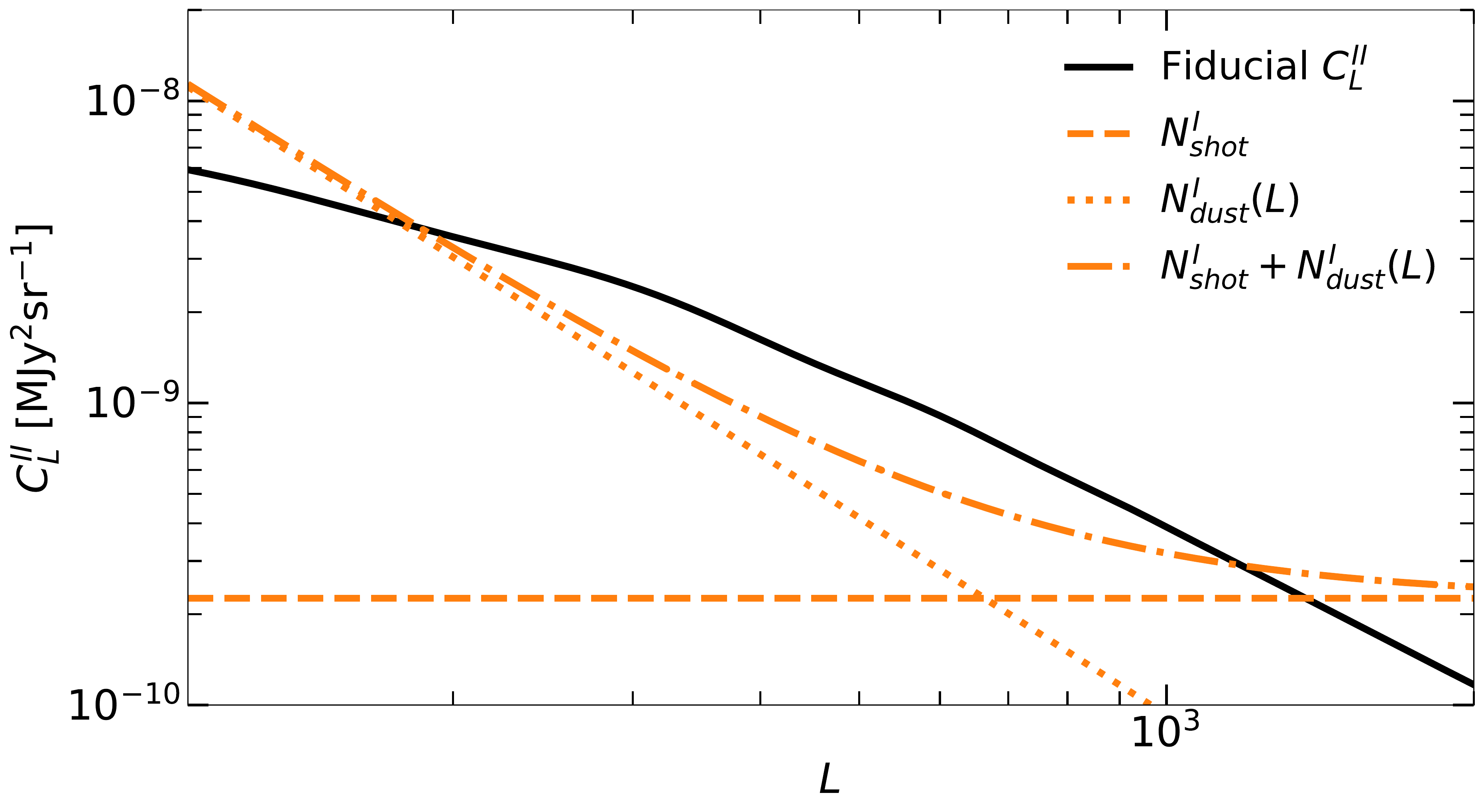}
	\caption{\small{The main noise contributions to Planck-like Cosmic Infrared Background (CIB) measurements are shown compared to the best-fit spectrum at 353GHz (solid line). Shot noise (dashed line) and dust noise (dotted line) are individually plotted alongside the total noise (dot-dashed line). All curves here have units of MJy$^2$sr$^{-1}$.
}}
	\label{fig:cib_ps}
\end{figure}

The CIB radiation is primarily emission from dust in galaxies.  Thus, similar to the galaxy density observable, it is also a biased tracer of the LSS. Using the single spectral energy distribution (SED) model \cite[]{Hall:2009rv}, the window function at frequency, $\nu$, is defined as
\begin{equation}\label{cib_window}
   W^{\nu}_I(\chi)=A_{\nu}\frac{\chi^2}{\left(1+z(\chi)\right)^2}\exp\left(-\frac{(z-z_c)^2}{2\sigma_z^2}\right)f_{\nu(1+z)},
\end{equation}
for
\begin{equation}
   f_{\nu}=
    \begin{cases}
        \left[\exp\left(\frac{h\nu}{k_BT}\right)-1\right]^{-1}\nu^{\beta+3} & (\nu \leqslant\nu')\\
        \left[\exp\left(\frac{h\nu}{k_BT}\right)-1\right]^{-1}\nu^{\beta+3}\left(\frac{\nu}{\nu'}\right)^{-\alpha} & (\nu >\nu')
    \end{cases},
\end{equation}
with $z_c=2$, $\sigma_z=2$, $\beta=2$, and $\alpha=2$.

There are two main components to the noise for the CIB
\begin{equation}
    N^{I,\nu}_L = N^{I,\nu}_{\mathrm{shot}} + N^{I,\nu}_{\mathrm{dust}}(L),
\end{equation}
from Poisson noise, and contamination from galactic dust emission. This assumes the CMB power spectra has been accurately removed, a safe assumption given the precision of recent CMB experiments \cite[]{PCP2018}. We do not account for instrumental noise which is expected to dominate on small scales. Instead, we set $L_{\textrm{max}}=2000$ for $C^{II}_L$. Similarly, due to the difficulty in optimally removing galactic dust contamination from the CIB maps, noise from dust dominates on large scales. We set $L_{\textrm{min}}=110$ to account for this. Above $L_{\textrm{min}}$, the dust is modelled by a simple two parameter power law
\begin{equation}
   N^{I,\nu}_{\mathrm{dust}}(L)=\beta_{\nu}L^{-\alpha_{\nu}}.
\end{equation}
Following \cite[]{Namikawa:2021gyh,Yu:2017djs}, the Generalized Needlet Internal Linear Combination (GNILC) algorithm \cite[]{Remazeilles:2011ze} is applied to Planck data \cite[]{Aghanim:2016pcc}, and $C^{II,\nu}_{L} + N^{I,\nu}_L$ is fit to the filtered map to obtain values for $A_{\nu}$, $\beta_{\nu}$, and $\alpha_{\nu}$.

The total CIB power spectrum at $\nu=353$ GHz is shown in Fig. \ref{fig:cib_ps} along with the noise contributions from sampling and dust for the multipole range $L_{\textrm{min}}<L<L_{\textrm{max}}$.

\begin{table}[]\centering
\begin{tabular}{|l|l|}
    \hline
 $XY$ & $f^{\phi}_{XY}(\boldsymbol{l},\boldsymbol{L})$\\ \hline \hline
$TT$ & $\boldsymbol{l}\cdot\boldsymbol{L}\tilde{C}^{T\nabla T}_{l} + \boldsymbol{l'}\cdot\boldsymbol{L}\tilde{C}^{T\nabla T}_{l'}$\\
$EE$ & $\left(\boldsymbol{l}\cdot\boldsymbol{L}\tilde{C}^{E\nabla E}_{l} + \boldsymbol{l'}\cdot\boldsymbol{L}\tilde{C}^{E\nabla E}_{l'}\right)h_E(\boldsymbol{l},\boldsymbol{l}')$\\
$EB$ & $\left(\boldsymbol{l}\cdot\boldsymbol{L}\tilde{C}^{E\nabla E}_{l} + \boldsymbol{l'}\cdot\boldsymbol{L}\tilde{C}^{B\nabla B}_{l'}\right)h_B(\boldsymbol{l},\boldsymbol{l}')$\\
$TE$ & $\boldsymbol{l}\cdot\boldsymbol{L}\tilde{C}^{T\nabla E}_{l}h_E(\boldsymbol{l},\boldsymbol{l}') + \boldsymbol{l'}\cdot\boldsymbol{L}\tilde{C}^{T\nabla E}_{l'}$\\
$TB$ & $\boldsymbol{l}\cdot\boldsymbol{L}\tilde{C}^{T\nabla E}_{l}h_B(\boldsymbol{l},\boldsymbol{l}')$\\ \hline
\end{tabular}
    \caption{\small{The lensing potential response functions for individual pairs of CMB maps $X$ and $Y$ required for lensing reconstruction. The geometric factors are defined as $h_E(\boldsymbol{l_1},\boldsymbol{l_2})=\cos(2(\theta_{l_1}-\theta_{l_2}))$, and $h_B(\boldsymbol{l_1},\boldsymbol{l_2})=\sin(2(\theta_{l_1}-\theta_{l_2}))$, and $\boldsymbol{l}'=\boldsymbol{L}-\boldsymbol{l}$. The non-perturbative response functions given here (slightly generalizing previous results \cite{Hu:2001kj,Maniyar:2021msb}) use the lensed gradient spectra defined in Ref.~\cite[]{Lewis:2011fk}. The BB response function is excluded along with the curl-like terms, $\tilde{C}^{XP_{\perp}}$, as they are subdominant relative to the other terms \cite[]{Hanson:2010rp}.}}\label{tab:resps_kappa}
\end{table}

\begin{figure}[t]
 	\includegraphics[width=\linewidth]{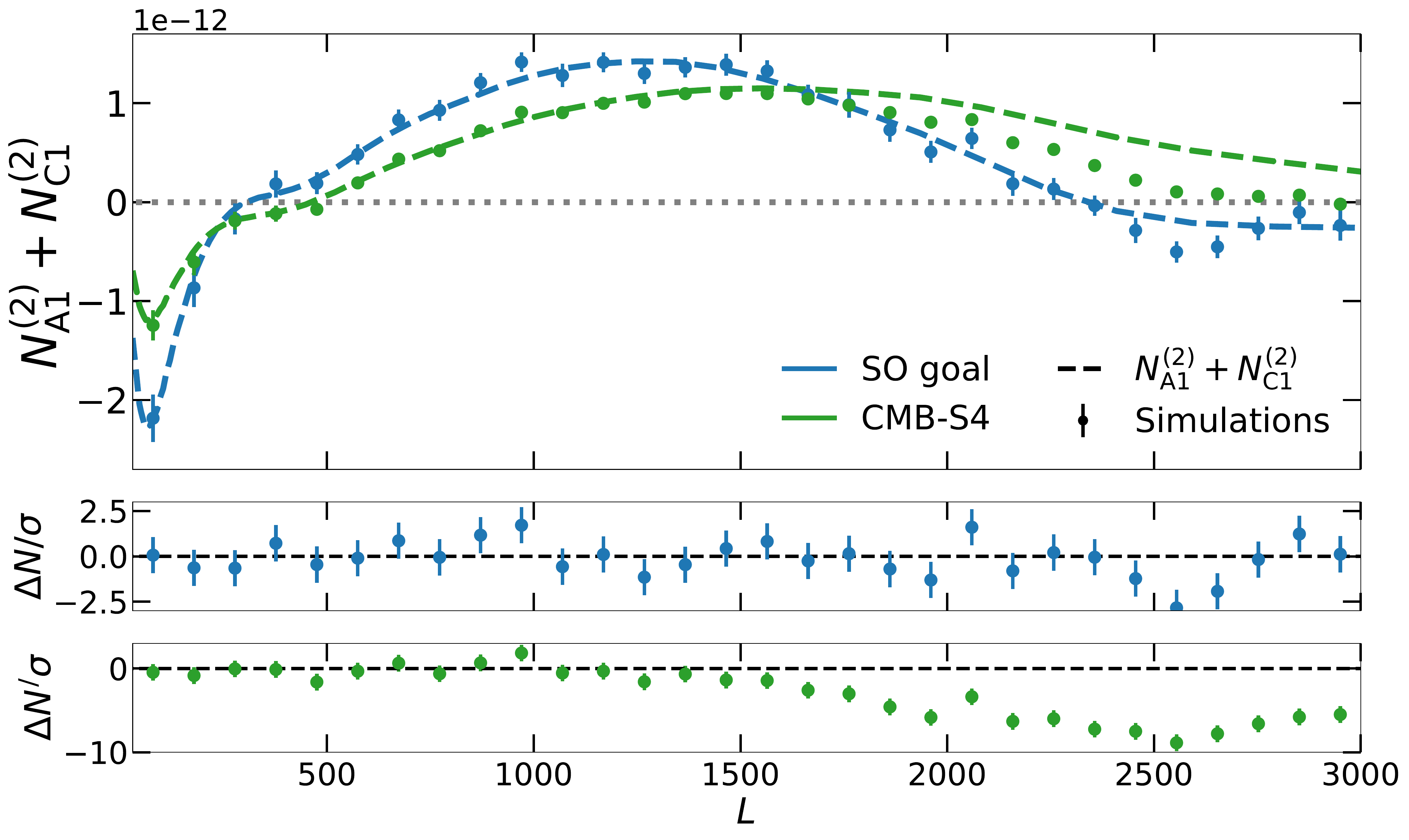}
	\caption{\small{The total Gaussian bias on the optimal rotation-template cross-spectrum $C^{\hat{\omega}\hat{\omega}^{\{\kappa,g,I\}}}$ for SO at goal sensitivity (blue) and CMB-S4 (green), assuming there are no contributions from contractions entering via the $\hat\kappa$ quadratic estimator. Therefore, only the combination $N^{(2)}_{\textrm{A1}}+N^{(2)}_{\textrm{C1}}$ (dashed lines) should be present. $\hat{\omega}$ and $\hat{\kappa}$ are reconstructed with the GMV estimator. Points represent the mean bias from 100 Monte Carlo simulations for 30 linearly-spaced bins in $L$, with error bars representing standard error in the mean. {\it Bottom panels:} Residuals of the total simulated Gaussian bias normalized by their 1-$\sigma$ errors.}}\label{fig:N2_bias}
\end{figure}

\begin{figure*}[t]
 	\includegraphics[width=\linewidth]{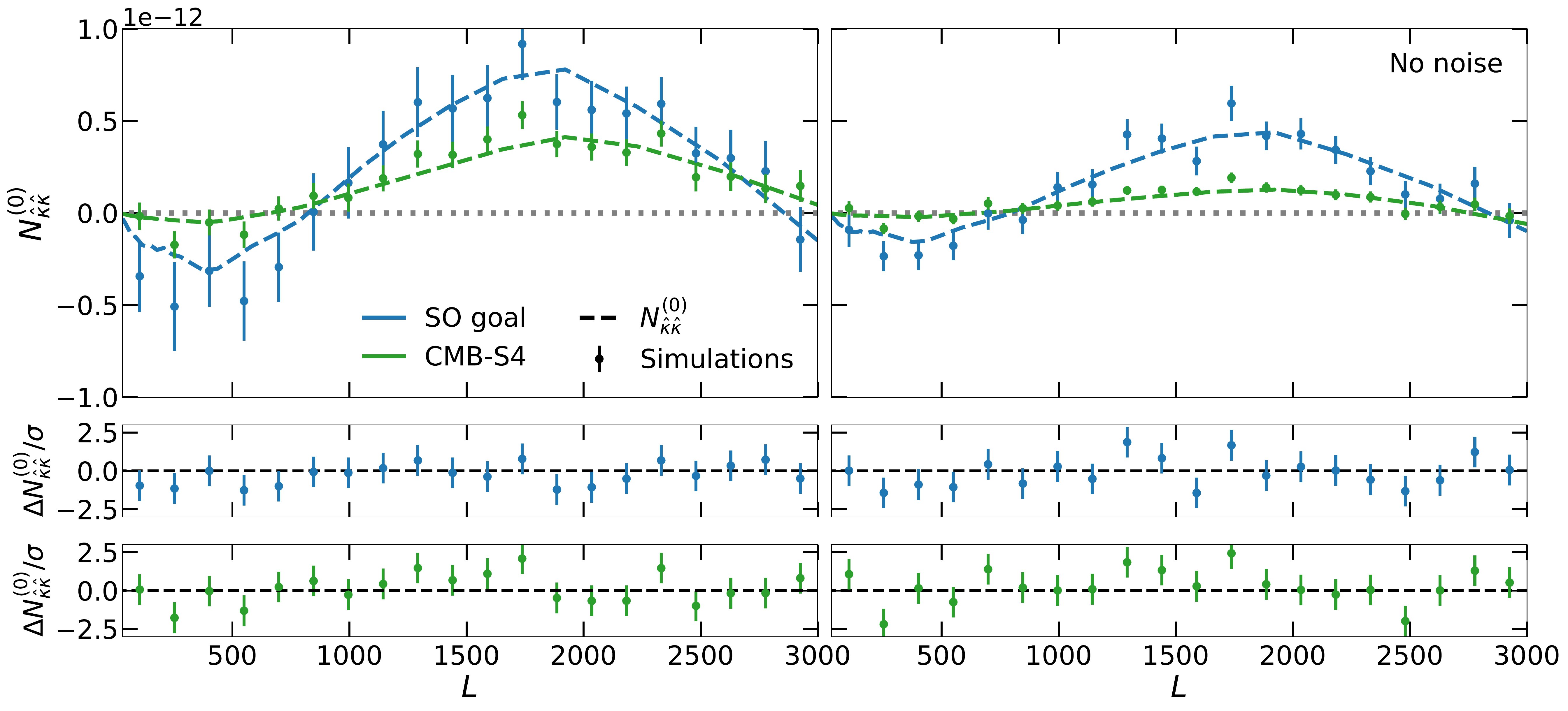}
	\caption{\small{The total $N^{(0)}$ bias on the optimal rotation-template cross-spectrum $C^{\hat{\omega}\hat{\omega}^{\{\kappa,g,I\}}}$ for SO at goal sensitivity (blue) and CMB-S4 (green), assuming $\hat{\omega}$ and $\hat{\kappa}$ are reconstructed with the GMV estimator. Dashed lines show the analytic predictions for $N^{(0)}_{\hat{\kappa}\hat{\kappa}}$. Points show the mean bias from 100 Monte Carlo simulations for 20 linearly-spaced bins in $L$, with error bars representing standard error in the mean. {\it Left panel:} Contributions to the bias from CMB map level noise are included in the theoretical curves and simulated results. This represents the case in which $\hat{\omega}$ and $\hat{\kappa}$ are reconstructed from the same CMB maps. {\it Right panel:} Same as left panel with exception that $\hat{\omega}$ and $\hat{\kappa}$ are now measured using CMB maps with independent noise. Therefore noise is not included in the simulations at map level, and similarly noise contributions are not included in the theory lines. {\it Bottom panels:} Residuals of the total simulated Gaussian bias normalized by their 1-$\sigma$ errors show good agreement between theroy and simulations.}}\label{fig:N0_bias_big}
\end{figure*}

\section{Isolating cross-spectrum biases}\label{app:bias}
Here, Gaussian Monte Carlo simulations are used to test the analytically derived results for each source of bias considered in this paper.
\subsection{$N^{(2)}_{\textrm{A1}} + N^{(2)}_{\textrm{C1}}$}
The $N^{(2)}$ biases are tested with Monte Carlo simulations using slight modifications to the methodology presented in Section \ref{sec:sims}. The first modification is not adding noise to any of the CMB or LSS maps. Noise does not contribute to either $N^{(2)}_{\textrm{A1}}$ or $N^{(2)}_{\textrm{C1}}$; including noise would only increase the scatter of the results. However, noise power is included in the filters used for the GMV reconstruction, i.e. in the inverse covariance weights of Eq.~\eqref{eq:gmv_weight}, and the inverse filtered LSS maps in Eq.~\eqref{eq:flat_omega_QE_explicit}.

The other modification is not using a lensing reconstructed convergence as one of the LSS tracer maps. Instead, we directly take the input $\kappa$ realization that performs the lensing on the {\sc LensIt} CMB maps as one of the inputs to the template QE. This ensures that the additional bias terms described in Section \ref{sec:bias_add} from contractions within the quadratic estimation of $\hat{\kappa}$ do not exist within the simulations. This methodology does not actually isolate $N^{(2)}_{\textrm{A1}}$ or $N^{(2)}_{\textrm{C1}}$, instead it captures the total Gaussian bias in the absence of these additional $\hat{\kappa}$ induced terms. This will include contributions from biases such as the $\mathcal{O}(\Psi^6)$ terms from Eq.~\eqref{eq:bias_expand}.

The top panel of Fig.~\ref{fig:N2_bias} shows the resulting total bias from 100 simulations overlaid with the analytical predictions. The residuals normalized with respect to the 1-$\sigma$ error in the bottom panel demonstrate good agreement that the total bias in this setup is well described by $N^{(2)}_{\textrm{A1}} + N^{(2)}_{\textrm{C1}}$. There is a slight discrepancy with the CMB-S4 results at $L>1500$, suggesting possible contributions from biases above $N^{(2)}$ level are getting picked up in the simulations. CMB-S4 would be more susceptible to such terms because of better lensing reconstruction on small scales (due to lower CMB noise, particularly in the polarization maps).

\begin{figure*}[t]
 	\includegraphics[width=\linewidth]{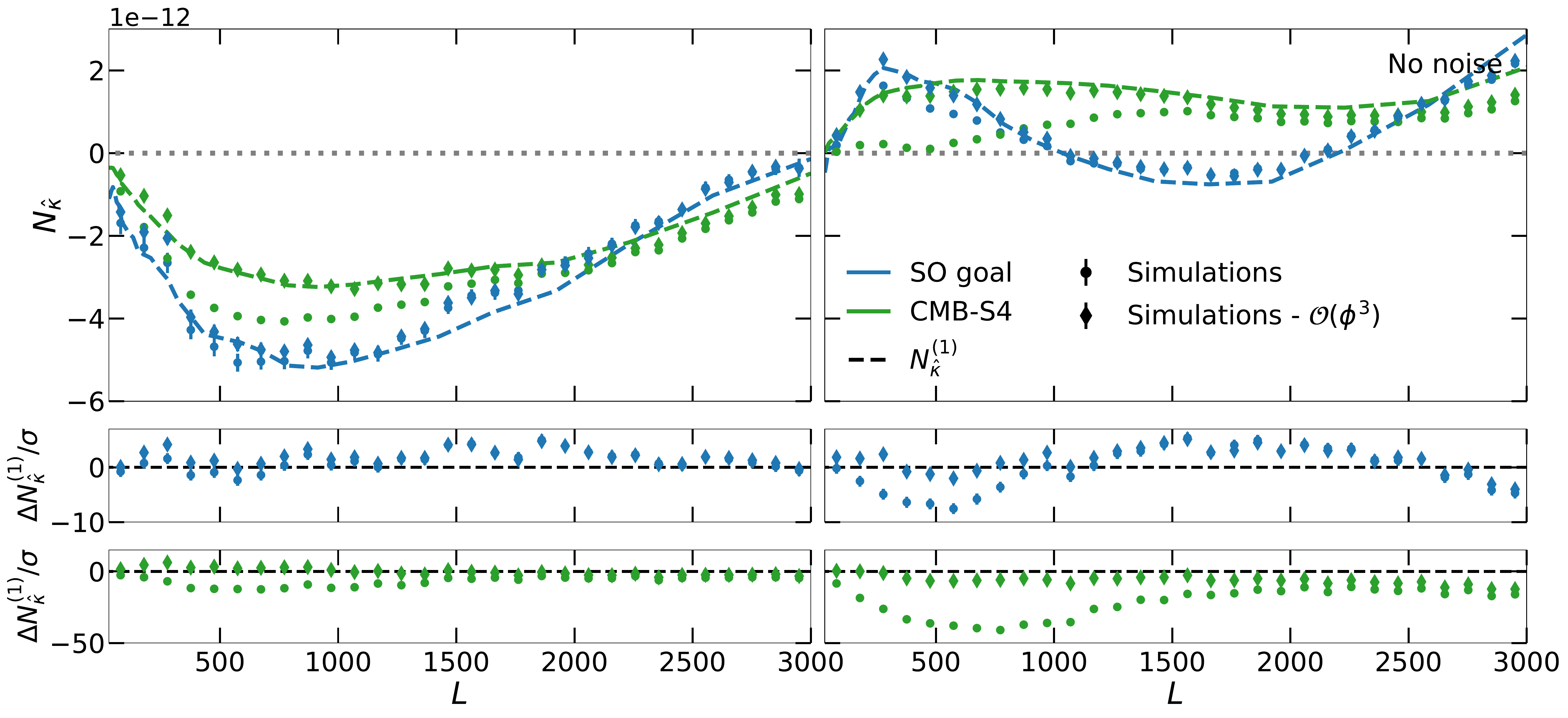}
	\caption{\small{Isolation of the one-$\kappa$ bias terms, with leading contribution from $N^{(1)}_{\hat{\kappa}}$, on the optimal rotation-template cross-spectrum $C^{\hat{\omega}\hat{\omega}^{\{\kappa,g,I\}}}$ for SO at goal sensitivity (blue) and CMB-S4 (green) assuming $\hat{\omega}$ and $\hat{\kappa}$ are reconstructed with the GMV estimator. Dashed lines show the analytic predictions for $N^{(1)}_{\hat{\kappa}}$. Points show the mean bias from 100 Monte Carlo simulations for 30 linearly-spaced bins in $L$, with error bars representing standard error in the mean. The diamond points are from the same simulation but with contributions that scale as $N^{(2)}_{\hat{\kappa}}$ subtracted, i.e. subtracting terms that scale as $\mathcal{O}(\phi^3\times{\rm tracer})$ estimated using a scaled simulation. {\it Left panel:} Contributions to the bias from CMB map level noise are included in the theoretical curves and simulated results. This represents the case in which $\hat{\omega}$ and $\hat{\kappa}$ are reconstructed from the same CMB maps. {\it Right panel:} Same as left panel with exception that $\hat{\omega}$ and $\hat{\kappa}$ are now measured using CMB maps with independent noise. Therefore noise is not included in the simulations at map level, and similarly noise contributions are not included in the theory lines. {\it Bottom panels:} Residuals of the total simulated Gaussian bias normalized by their 1-$\sigma$ errors (dots) show qualitative agreement between theory and simulations for SO goal. Residuals of the simulated bias with $N^{(2)}_{\hat{\kappa}}$ removed (diamonds) reduce the CMB-S4 discrepancy, and improve SO goal results at low $L$.}} \label{fig:N1_bias_big}
\end{figure*}

\subsection{$N^{(0)}_{\hat{\kappa}\hat{\kappa}}$}
To isolate all $N^{(0)}$ bias from simulations we use the same methodology as Section \ref{sec:sims} with one modification. The CMB maps used to ``reconstruct" $\hat{\omega}$ and $\hat{\kappa}$ are now Gaussian realizations of lensed CMB power-spectra. As the maps have not been deflected they therefore contain no lensing signal, while retaining the same correlations between maps. The resulting simulated rotation cross-spectrum can only contain bias signals at $N^{(0)}$ level as $\langle\phi a\rangle$ correlations in e.g. Eq.~\eqref{eq:mixed_bi} and Eq.~\eqref{eq:N1_alpha} (or $\langle\phi\phi\rangle$ in higher order terms) vanish.

We carry out the simulations for two cases; the first is with noise included in all maps, and the second is without noise. The results of these two cases are displayed in the left and right panels of Fig.~\ref{fig:N0_bias_big}. The total $N^{(0)}$ biases from 100 simulations are overlaid with the analytical predictions for $N^{(0)}_{\hat{\kappa}\hat{\kappa}}$. The residuals normalized with respect to the 1-$\sigma$ error in the bottom panels demonstrate good agreement that the total bias in both setups are well described by $N^{(0)}_{\hat{\kappa}\hat{\kappa}}$.

\subsection{$N^{(1)}_{\hat{\kappa}}$}\label{app:N1}
To test $N^{(1)}_{\hat{\kappa}}$ with simulations we follow a similar procedure to Refs.~\cite{PL2018,SPT:2014puc}. For each set of LSS simulations, we generate 2 independent realizations of unlensed CMB maps, $X^1\in\{T^1,E^1,B^1\}$ and $X^2\in\{T^2,E^2,B^2\}$, and use {\sc LensIt} to lens each set such that all 6 CMB maps are lensed with the same realization $\phi=\phi_1$. Then the lensing observables are quadratically estimated with the usual GMV estimator (Eq.~\eqref{eq:omega_qe} for $\hat{\omega}$), but in this case each QE leg contains a map from a different set, e.g. in Eq.~\eqref{eq:omega_qe} $\tilde{X}^i\in \tilde{X}^1_{\phi_1}$ and $\tilde{X}^j\in \tilde{X}^2_{\phi_1}$. Explicitly, we reconstruct the lensing observables such that

\begin{equation}\label{eq:omega12}
\hat{\omega}^{X^1,X^2}\equiv\frac{\hat{\omega}\left[\tilde{X}^1_{\phi_1},\tilde{X}^2_{\phi_1}\right]+\hat{\omega}\left[\tilde{X}^2_{\phi_1},\tilde{X}^1_{\phi_1}\right]}{2},
\end{equation}
and
\begin{equation}\label{eq:kappa12}
\hat{\kappa}^{X^1,X^2}\equiv\frac{\hat{\kappa}\left[\tilde{X}^1_{\phi_1},\tilde{X}^2_{\phi_1}\right]+\hat{\kappa}\left[\tilde{X}^2_{\phi_1},\tilde{X}^1_{\phi_1}\right]}{2}.
\end{equation}
The same maps are provided as input to $\hat{\omega}^{X^1,X^2}$ and $\hat{\kappa}^{X^1,X^2}$. This guarantees the independence between pairs of lensed CMB fields within the same QE, hence $N^{(2)}_{\textrm{A1}}$, $N^{(2)}_{\textrm{C1}}$, and $N^{(0)}_{\hat{\kappa}\hat{\kappa}}$ vanish. However, CMB cross-spectrum signals still exist from contractions between the $\hat{\omega}^{X^1,X^2}$ and $\hat{\kappa}^{X^1,X^2}$ estimators, so $N^{(1)}_{\hat{\kappa}}$ like signals still get picked up by the simulation. From the definitions Eq.~\eqref{eq:omega12} and Eq.~\eqref{eq:kappa12}, the simulations under-predict $N^{(1)}_{\hat{\kappa}}$ by a factor of two, this is accounted for by doubling the final simulated output.

Another consequence of this method is the $\hat{a}^i\hat{a}^j=\hat{\kappa}\hat{\kappa}$ term in Eq.~\eqref{eq:flat_omega_QE_explicit} always vanishes, therefore $N^{(1)}_{\hat{\kappa}\hat{\kappa}}$ biases vanish. However, the same is also true for the $N^{(1)}_{\hat{\kappa}\kappa}$ like terms that should be present in Eq.~\eqref{eq:N1_add}. To account for this, the $N^{(1)}_{\hat{\kappa}}$ predictions in Fig.~\ref{fig:N1_bias_big} do not include the $\hat{\kappa}\hat{\kappa}$ term in the numerical integration of Eq.~\eqref{eq:N1_add}.

We again perform the simulations for two cases; the first with noise included in all maps, and the second without noise. The results are displayed by the dots in Fig.~\ref{fig:N1_bias_big} for 100 simulations. There is decent agreement between simulation and theory $N^{(1)}_{\hat{\kappa}}$ for SO goal (blue dashed lines) only. The CMB-S4 simulated bias is discrepant compared to the analytic curves (green dashed lines) with and without noise, similar to the total bias results in Fig.~\ref{fig:bias_final_big}.

The disagreement between simulations and $N^{(1)}_{\hat{\kappa}}$ for CMB-S4 hints at missing contributions to the bias. Consider again the arguments for the leading one-$\kappa$ bias term, we have $\langle\hat{\omega}\hat{\omega}^{\textrm{tem}}\rangle\propto\langle\tilde{X}\tilde{Y}\tilde{X}\tilde{Y}\hat{a}^j\rangle$. Now expanding out the CMB fields using Eq.~\eqref{eq:lensed_field_expansion} we find next-to-leading order contractions such as $\langle \delta_{\phi}X\tilde{Y}\delta_{\phi}X\delta_{\phi}Y\hat{a}^j\rangle$, $\langle \delta_{\phi}^2X\tilde{Y}\tilde{X}\delta_{\phi}Y\hat{a}^j\rangle$, $\langle \delta_{\phi}^3X\tilde{Y}\tilde{X}\tilde{Y}\hat{a}^j\rangle$, etc at $N^{(2)}$ level. These terms form part of the $N^{(2)}_{\hat{\kappa}}$ bias, and are not cancelled by the simulated configuration in Fig.~\ref{fig:N1_bias_big}. To verify that the discrepancy can be attributed to $N^{(2)}_{\hat{\kappa}}$ we perform a simple scaling test. We re-compute the simulations with $\phi_1\rightarrow2\phi_1$, and find that the resulting offset increased by a factor of $\sim8$ as expected if $\Delta N\propto N^{(2)}_{\hat{\kappa}}$ (from contractions of three $\phi$ with one external tracer). The offset is then rescaled and subtracted from the total simulated bias, producing the diamond points in Fig.~\ref{fig:N1_bias_big}. This provides better agreement with theory $N^{(1)}_{\hat{\kappa}}$ and also improves SO goal at $L<1000$. Therefore, consideration of $N^{(2)}_{\hat{\kappa}}$ bias should also resolve the issues in Fig.~\ref{fig:bias_final_big} for CMB-S4. Derivation and further validation of $N^{(2)}_{\hat{\kappa}}$ is left to future work, but the full total bias can easily be estimated from simulations if the data can be simulated reliably.

\providecommand{\aj}{Astron. J. }\providecommand{\apj}{ApJ
  }\providecommand{\apjl}{ApJ
  }\providecommand{\mnras}{MNRAS}\providecommand{\prl}{PRL}\providecommand{\prd}{PRD}\providecommand{\jcap}{JCAP}\providecommand{\aap}{A\&A}

\end{document}